\documentclass[a4paper,11pt]{article}

\usepackage{jheppub} 

\usepackage{mathrsfs}
\usepackage{bbm}
\usepackage{bm}
\usepackage{graphicx}
\usepackage{epsfig}
\usepackage{graphicx,subfigure}
\usepackage[numbers,sort&compress]{natbib}
\usepackage{natbib}
\usepackage{array,longtable}
\usepackage{multirow}
\usepackage{slashed}
\usepackage[utf8]{inputenc}
\usepackage{arydshln}
\usepackage{float}
\usepackage{adjustbox}
\usepackage{xcolor}
\usepackage{soul}
\allowdisplaybreaks

\title{\bf \boldmath Study of $\tau^- \to \omega \pi^- \nu_\tau$ decay in resonance chiral theory with tensor sources}

\author[a]{Feng-Zhi Chen,}
\author[b,c,1]{Xin-Qiang Li,\note{Corresponding author.}}
\author[b]{Shi-Can Peng,}
\author[b,d]{Ya-Dong Yang}
\author[b]{{\small and} Yuan-He Zou}

\affiliation[a]{Department of Physics and Siyuan Laboratory, College of Physics $\&$ Optoelectronic Engineering, Jinan University, Guangzhou 510632, P.R. China}
\affiliation[b]{Institute of Particle Physics and Key Laboratory of Quark and Lepton Physics~(MOE), Central China Normal University, Wuhan, Hubei 430079, China}
\affiliation[c]{Center for High Energy Physics, Peking University, Beijing 100871, China}
\affiliation[d]{Institute of Particle and Nuclear Physics, Henan Normal University, Xinxiang, Henan 453007, China}

\emailAdd{fzchen@jnu.edu.cn}
\emailAdd{xqli@mail.ccnu.edu.cn}
\emailAdd{shicanpeng@mails.ccnu.edu.cn}
\emailAdd{yangyd@mail.ccnu.edu.cn}
\emailAdd{yuanhe99zou@mails.ccnu.edu.cn}

\abstract{In this work, we make a study of the $\tau^- \to \omega\pi^-\nu_\tau$ decay in the framework of low-energy effective field theory. The $J^{\mathcal{P}G}$ decompositions of the quark currents and the $\omega\pi$ final state show that, besides the Standard Model vector interaction, only the non-standard tensor interaction can have a non-zero contribution to the decay. To discuss its effect, a reliable calculation of the $\omega\pi$ tensor form factors is necessary. After constructing the Lagrangian of resonance chiral theory with external tensor sources, we calculate both the vector and tensor form factors with the relevant resonance couplings determined by combining the QCD short-distance constraints, the fit to the spectral function of $\tau^- \to \omega\pi^-\nu_\tau$ decay, as well as the matching between the $\mathcal{O}(p^4)$ odd-intrinsic-parity operators after integrating out the vector resonances and the $\mathcal{O}(p^6)$ operators of chiral perturbation theory. The new physics effect is then investigated in the distributions of the spectral function and the forward-backward asymmetry of $\tau^- \to \omega\pi^-\nu_\tau$ decay. We find that the spectral function is dominated by the Standard Model, and the non-standard tensor contribution is negligible. However, since the forward-backward asymmetry can be only generated with a non-zero tensor interaction, the observable is quite sensitive to this kind of new physics. A future measurement of the observable at the Belle II experiment as well as at the proposed Tera-Z and STCF facilities is, therefore, strongly called for to check the existence of such a non-standard tensor interaction.}

\begin{document} 
\maketitle
\flushbottom

\section{Introduction}
\label{sec:intro}

The hadronic tau decays play an important role not only in the determination of the quantum chromodynamics (QCD) coupling constant $\alpha_s(m_\tau)$ and the extraction of the Cabibbo-Kobayashi-Maskawa (CKM)~\cite{Cabibbo:1963yz,Kobayashi:1973fv} quark mixing matrix element $V_{us}$ in the Standard Model (SM)~\cite{ALEPH:2005qgp,Davier:2005xq,Pich:2013lsa,Pich:2020gzz,dEnterria:2022hzv}, but also in the probe of new physics (NP) beyond the SM~\cite{Cirigliano:2018dyk,Gonzalez-Solis:2020jlh,Cirigliano:2021yto,Chen:2021udz}. To have a sensitive constraint on the latter, one should use the most precise data on hadronic tau decays and, at the same time, provide a reliable description of the hadronic matrix elements with well-controlled theoretical uncertainty. Although the hadronic tau decays into one, two, and three pseudo-scalar mesons with light-quark content (namely $u$, $d$ and $s$) have been extensively studied both within the SM and beyond, there are few dedicated investigations of the decays with one vector ($V$) and one pseudo-scalar ($P$) mesons in the final state,  denoted generically by $\tau\to VP\nu_\tau$ with $V=\rho$, $K^\ast$, $\omega$, $\phi$ and $P=\pi$, $K$, $\eta^{(\prime)}$. Most of the previous studies focus mainly on the (differential) branching ratios, the hadron spectral functions, and the forward-backward asymmetries of $\tau\to VP\nu_\tau$ decays within the SM, with the hadronic form factors calculated in the vector meson dominance (VMD) model~\cite{Decker:1992jy,LopezCastro:1996xh,Flores-Tlalpa:2007hbv}, the extended Nambu-Jona-Lasinio model~\cite{Volkov:2012gv,Ahmadov:2015oca,Volkov:2019cja,Volkov:2019yli}, as well as the chiral perturbation theory ($\chi$PT)~\cite{Davoudiasl:1995ed,Nasriddinov:1998ag,Fajfer:1992hi} and resonance chiral theory (R$\chi$T)~\cite{Guo:2008sh}. Explorations of possible NP effects in these decays are, however, almost missing. This may attribute to the lack of reliable description of the hadronizations of light-quark bilinears in the presence of non-standard interactions, \textit{i.e.}, some novel form factors needed to describe the decays. To this end, a good knowledge of these form factors is strongly called for.

In this work, we attempt to make a preliminary study of the NP effect in one of the aforementioned tau decay modes, $\tau^-\to\omega\pi^-\nu_\tau$, which is usually employed to test the existence of the so-called second-class current (SCC)~\cite{Weinberg:1958ut,Leroy:1977pq,Berger:1987ku}. According to its definition, the SCC has an opposite $G$-parity to that of the first-class current (FCC): the $J^{\mathcal{P}G}$ quantum numbers are $0^{++}$, $0^{--}$, $1^{+-}$, $1^{-+}$ for the FCC, while they are $0^{+-}$, $0^{-+}$, $1^{++}$, $1^{--}$ for the SCC.\footnote{Here $G\equiv \mathcal{C}e^{i\pi I_2}$ with $\mathcal{C}$ being the charge conjugation operator and $I_2$ the second generator of isospin rotations~\cite{Lee:1956sw}. $J$ and $\mathcal{P}$ denote the total angular momentum and parity, respectively.} In the SM, the $\tau^-\to\omega\pi^-\nu_\tau$ decay is dominated by the FCC, while the SCC contribution can only arise due to the $G$-parity violating effect and is, therefore, highly suppressed with respect to that of the FCC. Experimentally, the total branching ratio of $\tau^-\to\omega\pi^-\nu_\tau$ decay has been measured to be $(1.95\pm0.06)\%$~\cite{ParticleDataGroup:2022pth}, while the current experimental upper limit for the SCC contribution is only of $1.4\times10^{-4}$ at $90\%$ confidence level (CL)~\cite{BaBar:2009jyj}. On the theoretical aspect, the SM estimate of this decay based on the VMD model gives a branching ratio of $(1.5\sim2.8)\times10^{-5}$ for the isospin-breaking induced SCC effect~\cite{Paver:2012tq}. This implies that there still leaves a large room for the existence of genuine SCC induced by non-standard interactions. However, as we can see from these numbers, even there may exist genuine SCC contribution in this decay, its upper limit is two orders of magnitude smaller than that of the total branching ratio (more precisely, the former is even four times smaller than the uncertainty of the latter). Thus, we can safely neglect the SCC contribution for the moment. Furthermore, it is not possible to separate the SCC contribution from that of the FCC, unless the full knowledge of the form factors is known and the measurements of the angular observables of this decay are precise enough. Therefore, in this work, we shall assume that the $\tau^-\to\omega\pi^-\nu_\tau$ decay is dominated by the FCC non-standard interactions, while the $G$-parity is assumed to be strictly conserved.

To study the NP effects in a model-independent way, we shall work in the framework of a low-energy effective field theory (LEFT)~\cite{Jenkins:2017jig,Jenkins:2017dyc}, which is invariant under the Lorentz transformation and respects the $SU(3)_C \otimes U(1)_{em}$ gauge symmetry. The only left degrees of freedom in LEFT are the light quarks ($u$, $d$, $s$), charged leptons ($e$, $\mu$, $\tau$), neutrinos ($\nu_e$, $\nu_\mu$, $\nu_\tau$), gluons, and photons, while the remaining particles of the SM are all integrated out and there are no any exotic NP particles with masses below $\sim 2~\text{GeV}$~\cite{Cirigliano:2009wk,Bhattacharya:2011qm}. In recent years, many hadronic tau decays have been studied in such a framework; see, \textit{e.g.}, Refs.~\cite{Garces:2017jpz,Cirigliano:2017tqn,Miranda:2018cpf,Cirigliano:2018dyk,Rendon:2019awg,Chen:2019vbr,Gonzalez-Solis:2019lze,Gonzalez-Solis:2020jlh,Chen:2020uxi,Arroyo-Urena:2021dfe,Arroyo-Urena:2021nil,Chen:2021udz,Cirigliano:2021yto,Arteaga:2022xxy}. Here we proceed to apply the same framework to study possible NP impacts on the $\tau^-\to\omega\pi^-\nu_\tau$ decay. After a detailed analysis of the $J^{\mathcal{P}G}$ quantum numbers of the $\omega\pi$ system as well as the quark-current operators present in the LEFT Lagrangian, one can see that, besides the SM vector contribution, only a non-standard tensor interaction can have an impact on the decay. To estimate the tensor contribution to the $\tau^-\to\omega\pi^-\nu_\tau$ decay, an explicit calculation of the $\omega\pi$ tensor form factors is necessary. As the $\omega\pi$ invariant mass $\sqrt{s}$ in the tau decay is ranged from the $\omega\pi$ threshold up to the tau mass, there are plenty of resonances populating the $\omega\pi$ mass spectrum in such a large kinematic range. Thus, we shall resort to a theory that incorporates both the lowest-lying pseudo-scalars and the relevant resonances. One of the avenues to realize this task is to calculate the $\omega\pi$ tensor form factors by employing the R$\chi$T, which was constructed firstly in Refs.~\cite{Ecker:1988te,Ecker:1989yg} and then developed to include the odd-intrinsic-parity interactions in Refs.~\cite{Ruiz-Femenia:2003jdx,Cirigliano:2004ue}. Let us recall that the vector and axial-vector form factors for a generic $\tau\to VP\nu_\tau$ decay have been worked out in Ref.~\cite{Guo:2008sh}, in the R$\chi$T framework. However, to derive the needed $VP$ tensor form factors, we have to extend the framework by constructing the chirally invariant Lagrangian with external tensor sources. The most general Lagrangian for the lowest-lying pseudo-scalars in the presence of external sources coupled to the tensor Dirac bilinear $\bar{\psi}\sigma_{\mu\nu}\psi$ has been worked out in Ref.~\cite{Cata:2007ns}, extending therefore the conventional chiral Lagrangian~\cite{Weinberg:1978kz,Gasser:1983yg,Gasser:1984gg}. In addition, the interactions between the vector resonances and the external tensor fields have been studied in, \textit{e.g.}, Refs.~\cite{Mateu:2007tr,Miranda:2018cpf,Chen:2019vbr,Husek:2020fru,Shi:2020rkz}. Nevertheless, the effective Lagrangian used to describe the interactions among the lowest-lying pseudo-scalars, the vector resonances, and the external tensor sources has not been presented yet, which will be therefore one of the main tasks of this work. Its explicit construction will be detailed in appendix~\ref{app:constructVTP}. In principle, the tensor resonances could also contribute to the $\omega\pi$ tensor form factors, and should be therefore added into the Lagrangian as explicit degrees of freedom as well. However, the VMD picture tells us that the vector resonances will dominate the total contribution, making therefore the contribution from the tensor resonances negligible. Thus, during the calculation of the $\omega\pi$ tensor form factors throughout this work, we shall only consider contributions from the vector resonances. Since the resonance couplings in the Lagrangian cannot be determined by the R$\chi$T itself, we shall resort to other constraints to obtain the required $\omega\pi$ tensor form factors. This can be achieved with the help of, \textit{e.g.}, the QCD short-distance constraints~\cite{Brodsky:1973kr,Lepage:1979zb,Lepage:1980fj}, the matching between the $\mathcal{O}(p^4)$ odd-intrinsic-parity R$\chi$T operators after integrating out the vector resonances and the $\mathcal{O}(p^6)$ $\chi$PT operators, as well as the fit to the measured spectral function of $\tau^- \to \omega \pi^- \nu_\tau$ decay~\cite{CLEO:1999heg}. Since our ultimate aim is to investigate the NP impacts on the $\tau^- \to \omega \pi^- \nu_\tau$ decay, with the $\omega\pi$ tensor form factors in hand, we shall further investigate the sensitivities of the spectral function and the forward-backward asymmetry of the decay to the non-standard tensor contribution. 

The rest of this paper is organized as follows. In section~\ref{sec:EFT}, we firstly describe the $\tau^- \to \omega \pi^- \nu_\tau$ decay in the LEFT framework, and then give the analytic expressions of the spectral function and forward-backward asymmetry of the decay in the presence of the non-standard tensor interaction. The calculation of the $\omega\pi$ vector and tensor form factors in the R$\chi$T framework is presented in section~\ref{sec:FFs}. Our numerical analysis and conclusions are finally given in sections~\ref{sec:numerical} and \ref{sec:conclusion}, respectively. For convenience, the construction of the R$\chi$T Lagrangian with external tensor sources and the determination of the resonance couplings present in the Lagrangian are relegated to appendices~\ref{app:constructVTP} and \ref{app:Birelations}, respectively.

\section{\boldmath $\tau^- \to \omega\pi^-\nu_\tau$ decay in LEFT}
\label{sec:EFT}

In this section, we demonstrate how to describe the $\tau^- \to \omega \pi^- \nu_\tau$ decay in the LEFT framework, and present the analytic expressions of the spectral function and the forward-backward asymmetry of the decay in the presence of a non-standard tensor interaction.  

\subsection[LEFT Lagrangian for $\tau^-\to \bar{u}d\nu_\tau$]{\boldmath LEFT Lagrangian for $\tau^-\to \bar{u}d\nu_\tau$}

The $\tau^- \to \omega\pi^-\nu_\tau$ decay is mediated by the quark-level $\tau^-\to \bar{u}d\nu_\tau$ transition. Without loss of generality, the NP impacts on this process can be described by a subset of the model-independent LEFT Lagrangian~\cite{Jenkins:2017jig,Jenkins:2017dyc}, which is valid below the electroweak scale and invariant under the Lorentz transformation and the $SU(3)_C\otimes U(1)_{em}$ gauge symmetry. In the absence of right-handed neutrinos and other light NP particles with masses below $\sim 2~\text{GeV}$, the most general LEFT Lagrangian describing the $\tau^-\to \bar{u}d\nu_\tau$ transition is given by~\cite{Cirigliano:2009wk,Bhattacharya:2011qm}\footnote{Note that the Dirac-algebra identity $\sigma^{\mu\nu}\gamma_5=-\frac{i}{2}\epsilon^{\mu\nu\alpha\beta}\sigma_{\alpha\beta}$ has been used to obtain the tensor interaction term in Eq.~\eqref{eq:LEFT}, where $\epsilon^{\mu\nu\alpha\beta}$ is the anti-symmetric Levi-Civita tensor, with the convention $\epsilon^{0123}=-1$. We have not included the wrong-flavor neutrino interactions, which do not interfere with the SM amplitude and hence contribute to the observables only at $\mathcal{O}(\hat{\epsilon}^2_X)$.} 
\begin{align}\label{eq:LEFT}
	\mathcal{L}_\mathrm{eff} = &-\frac{G_F^0 V_{ud}}{\sqrt{2}}  \left(1+\epsilon_L +\epsilon_R\right) \Big\{ \bar{\tau} \gamma_\mu (1-\gamma_5) \nu_\tau \cdot \left[ \bar{u} \gamma^\mu d -(1-2\hat{\epsilon}_R) \bar{u} \gamma^\mu \gamma_5 d \right] \nonumber \\
	& + \bar{\tau} (1-\gamma_5) \nu_\tau \cdot \left[ \hat{\epsilon}_S \bar{u} d -\hat{\epsilon}_P \bar{u} \gamma_5 d \right] + 2\hat{\epsilon}_T \bar{\tau} \sigma_{\mu\nu} (1-\gamma_5) \nu_\tau \cdot \bar{u} \sigma^{\mu\nu} d  \Big\} + \mathrm{h.c.}\,,
\end{align}
where $\sigma^{\mu\nu}=\frac{i}{2}[\gamma^\mu,\gamma^\nu]$, $G_F^0$ is the Fermi constant without NP contributions, and $V_{ud}$ the CKM matrix element involved in the transition. The hatted Wilson coefficients $\hat{\epsilon}_i=\epsilon_i / \left(1+\epsilon_L+\epsilon_R \right)$, with $i=R, S, P, T$, have been introduced to parameterize the different types of non-standard contributions normalized to the vector one, and the SM case is recovered when all $\epsilon_i=0$. As we are not concerned with the charge-parity violating observerbles in this work, all the $\epsilon_i$ are assumed to be real. Note also that these coefficients are renormalization scale and scheme dependent~\cite{Gonzalez-Alonso:2017iyc}, and all the numerical values quoted in this work are obtained at $\mu = 2~\text{GeV}$ in the modified minimal subtraction scheme. Such an LEFT framework has been widely adopted in the literature; see, \textit{e.g.}, Refs.~\cite{Garces:2017jpz,Cirigliano:2017tqn,Miranda:2018cpf,Cirigliano:2018dyk,Rendon:2019awg,Chen:2019vbr,Gonzalez-Solis:2019lze,Gonzalez-Solis:2020jlh,Chen:2020uxi,Arroyo-Urena:2021dfe,Arroyo-Urena:2021nil,Chen:2021udz,Cirigliano:2021yto}. 

\subsection[Decay amplitude and observables of $\tau^- \to \omega\pi^-\nu_\tau$]{\boldmath Decay amplitude and observables of $\tau^- \to \omega\pi^-\nu_\tau$}

For the $\tau^- \to \omega\pi^-\nu_\tau$ decay, not all the operators present in Eq.~\eqref{eq:LEFT} can provide a non-vanishing contribution, but only the ones with the same $J^{\mathcal{P}G}$ quantum numbers as of the $\omega\pi^-$ final state are relevant. Following a similar procedure as in Refs.~\cite{Ji:2000id,Papucci:2021pmj}, we can perform the angular momentum-parity ($J^\mathcal{P}$) decomposition of a given quark-current operator $\mathcal{O}\equiv\bar{d}\Gamma u$, where $\Gamma=1$, $\gamma_5$, $\gamma_\mu$, $\gamma_\mu\gamma_5$, and $\sigma_{\mu\nu}$ for scalar ($S$), pseudo-scalar ($P$), vector ($V$), axial-vector ($A$), and tensor ($T$) operators, respectively. Explicitly, we have
\begin{align} \label{eq:OJP}
J^\mathcal{P}(\mathcal{O}_S)=&0^+\,, &  J^\mathcal{P}(\mathcal{O}_P)=&0^-\,,\notag\\[0.1cm]
J^\mathcal{P}(\mathcal{O}_V)=&0^+\oplus 1^-\,, &  J^\mathcal{P}(\mathcal{O}_A)=&0^-\oplus 1^+\,,\notag\\[0.1cm]
J^\mathcal{P}(\mathcal{O}_T)=&1^+\oplus 1^-\,. &  &
\end{align}
Since the intrinsic spin-parity of $\omega$ and $\pi^-$ are $1^-$ and $0^-$ respectively, the $J^\mathcal{P}$ quantum number of the $\omega\pi^-$ system is given by $1^-\otimes0^-=1^+$. Then, for the $\omega\pi^-$ system with an orbital angular momentum $L=0, 1, 2, \cdots$ and a parity $(-1)^L$, one has the following states:
\begin{align}
J^\mathcal{P}(\omega\pi\mid_{L=0})&=1^+\otimes 0^+=1^+\,,\notag\\[0.1cm]
J^\mathcal{P}(\omega\pi\mid_{L=1})&=1^+\otimes 1^-=0^- \oplus 1^-\oplus 2^-\,,\notag\\[0.1cm]
J^\mathcal{P}(\omega\pi\mid_{L=2})&=1^+\otimes 2^+=1^+ \oplus 2^+\oplus 3^+\,.
\end{align}
Note that no partial wave with $L > 2$ contains $J \le 1$ and, to match with Eq.~\eqref{eq:OJP}, we need only consider the partial waves up to $J=1$~\cite{Papucci:2021pmj}. Furthermore, the intrinsic $G$-parity of both $\omega$ and $\pi^-$ is $-1$ and thus the $\omega\pi^-$ system has a $G$-parity of $+1$, while the $G$-parity of a given operator $\mathcal{O}$ can be determined from the following $G$-parity transformation:
\begin{align}
G\bar{d}\Gamma uG^{-1} =\,& G\bar{\psi}\Gamma\sigma_{-}\psi G^{-1}\notag\\
=&-\bar{\psi}\Gamma^\mathcal{C}\sigma_{-}\psi\,,
\end{align}
where $\psi=(u,d)^\top$, with $\top$ referring to the transpose operation, is the isospin doublet of the first-generation quarks, $\sigma_\pm=\frac{1}{2}(\sigma_1\pm i\sigma_2)$ with $\sigma_i$ being the Pauli matrices, and $\Gamma^\mathcal{C}=\mathcal{C}\Gamma \mathcal{C}^{-1}$ stands for the charge transformation of the Dirac matrix $\Gamma$. A direct calculation then yields
\begin{align}
\mathcal{C}\Big\{\mathcal{O}_S\,,\mathcal{O}_P\,,\mathcal{O}_V\,,\mathcal{O}_A\,,\mathcal{O}_T\Big\}\mathcal{C}^{-1}=\Big\{+\mathcal{O}_S\,,+\mathcal{O}_P\,,-\mathcal{O}_V\,,+\mathcal{O}_A\,,-\mathcal{O}_T\Big\}\,,
\end{align}
from which one immediately obtains 
\begin{align}
G\Big\{\mathcal{O}_S\,,\mathcal{O}_P\,,\mathcal{O}_V\,,\mathcal{O}_A\,,\mathcal{O}_T\Big\}G^{-1}=\Big\{-\mathcal{O}_S\,,-\mathcal{O}_P\,,+\mathcal{O}_V\,,-\mathcal{O}_A\,,+\mathcal{O}_T\Big\}\,.
\end{align}

As a summary, we collect in Table~\ref{tab:JPG} the $J^{\mathcal{P}G}$ quantum numbers of the quark-current operators as well as the $\omega\pi^-$ system with three different orbital angular momenta, from which one can see that only the two hadronic matrix elements, $\langle\omega\pi^-|\mathcal{O}_{V}|0\rangle$ and $\langle\omega\pi^-|\mathcal{O}_{T}|0\rangle$, are relevant for the $\tau^- \to \omega\pi^-\nu_\tau$ decay, where the operator $\mathcal{O}_V$ contributes in a $P$-wave ($1^{-+}$) manner and is dominated by the SM vector interaction, whereas $\mathcal{O}_T$ provides both an $S$-wave ($1^{++}$) and a $P$-wave amplitude. As discussed in section~\ref{sec:intro}, since the component $1^{++}$ corresponds to the SCC whose contribution is negligibly small, in the following we shall only focus on the contribution from the FCC component with $J^{\mathcal{P}G}=1^{-+}$.

\begin{table}[t]
  \tabcolsep 0.005in
  \renewcommand\arraystretch{1.5}   
  \begin{center}
  \begin{tabular}{|c|c|c|c|c|c|c|c|c|}
  \hline\hline
   & $\mathcal{O}_S$ & $\mathcal{O}_P$ & $\mathcal{O}_V$ & $\mathcal{O}_A$ & $\mathcal{O}_T$ & $\omega\pi\mid_{L=0}$& $\omega\pi\mid_{L=1}$& $\omega\pi\mid_{L=2}$\\
  \hline
  $J^{\mathcal{P}G}$ &  $0^{+-}$ & $0^{--}$ & $0^{++}\oplus 1^{-+}$ & $0^{--}\oplus 1^{+-}$ & $1^{++}\oplus 1^{-+}$& $1^{++}$ & $0^{-+}\oplus 1^{-+}\oplus 2^{-+}$ & $1^{++}\oplus2^{++}\oplus3^{++}$\\ 
  \hline \hline
  \end{tabular} 
  \caption{\small The $J^{\mathcal{P}G}$ of quark-current operators $\mathcal{O}_i$ for $i=S,P,V,A,T$, as well as of $\omega\pi$ system with orbital angular momentum $L=0,1,2$, respectively. \label{tab:JPG}}
  \end{center}  
\end{table}

After this analysis, we can therefore simply neglect those terms that do not contribute to the $\tau^-(p) \to \omega(p_2) \pi^-(p_1) \nu_\tau(p_3)$ decay, and write the invariant amplitude as
\begin{align}\label{eq:amplitude}
	\mathcal{M} = \mathcal{M}_V + \mathcal{M}_T = -i \frac{G_F^0 V_{ud}^\ast \sqrt{S_{EW}}}{\sqrt{2}} (1+\epsilon_L^\ast +\epsilon_R^\ast) \left[ L_\mu H^\mu + 2\hat{\epsilon}_T^\ast L_{\mu\nu} H^{\mu\nu} \right]\,,
\end{align}
where $S_{EW}=1.0201(3)$ stands for the short-distance electroweak radiative correction~\cite{Marciano:1985pd,Marciano:1988vm,Braaten:1990ef,Erler:2002mv},\footnote{Here the short-distance correction $S_{EW}$ is taken as a global factor, although it does not affect the tensor contribution. This approximation renders the expressions of various observables simpler while guaranteeing the resulting error to be negligibly small~\cite{Garces:2017jpz,Miranda:2018cpf}.} and $L_{i}$ and $H^{i}$ denote the leptonic currents and the hadronic matrix elements of the quark currents between the vacuum and the $\omega\pi^-$ final state respectively, with
\begin{align}
	L_\mu &= \bar{u}_{\nu_\tau}(p_3) \gamma_\mu (1-\gamma_5) u_\tau(p)\,,  \\
	L_{\mu\nu} &= \bar{u}_{\nu_\tau}(p_3) \sigma_{\mu\nu} (1+\gamma_5) u_\tau(p)\,,
\end{align} 
and
\begin{align}
\label{eq:Had1} 	
	H^\mu &=\langle \omega(p_2,\varepsilon)\pi^-(p_1) | \bar{d} \gamma^\mu u | 0 \rangle = i F_{V}(s) \epsilon^{\mu\nu\rho\sigma} \varepsilon_\nu^\ast p_{1\rho} p_{2\sigma}\,,  \\[0.2cm]
\label{eq:Had2} 
	H^{\mu\nu} &=\langle \omega(p_2,\varepsilon)\pi^-(p_1) | \bar{d} \sigma^{\mu\nu} u | 0 \rangle \notag \\
	&= F_{T1}(s) \epsilon^{\mu\nu\rho\sigma} (\varepsilon^\ast \cdot p_1) p_{1\rho} p_{2\sigma}- F_{T2}(s) \epsilon^{\mu\nu\rho\sigma} \varepsilon^\ast_\rho p_{1\sigma} - F_{T3}(s) \epsilon^{\mu\nu\rho\sigma} \varepsilon^\ast_\rho p_{2\sigma}\,.
\end{align}
Here $\varepsilon_\nu$ is the $\omega$ polarization four-vector that satisfies $p_2 \cdot \varepsilon=0$, $s=q^2=(p_1+p_2)^2$ is the $\omega \pi$ invariant mass squared, and $F_{V}(s)$ and $F_{Ti}(s)$ ($i=1,2,3$) denote the vector and tensor form factors, respectively. The explicit expressions of these form factors are listed in Eqs.~\eqref{eq:FV}--\eqref{eq:FT3}. Let us recall that in the VMD picture, the vector form factor $F_{V}(s)$ is dominated by the spin-1 resonance $\rho(770)$ and, to a less extent, by $\rho(1450)$, which in this work are abbreviated as $\rho$ and $\rho^\prime$, respectively. These vector resonances contribute also to the tensor form factors $F_{Ti}(s)$. To facilitate the calculations of these form factors, we shall adopt the anti-symmetric tensor field formalism to describe these vector resonances~\cite{Ecker:1988te,Ecker:1989yg}, in the framework of R$\chi$T; for more details, the readers are referred to section~\ref{sec:FFs}.

Working in the $\omega \pi$ rest frame and after integrating over the unobserved neutrino direction, we can write the doubly differential decay width of $\tau^-\to \omega \pi^- \nu_\tau$ as
\begin{align}\label{eq:dddw}
\frac{d^2\Gamma(\tau^- \to \omega \pi^- \nu_\tau)}{ds \, d\cos\theta}=A(s)+B(s)\cos\theta +C(s)\cos^2\theta\,,
\end{align}
where $\theta$ is defined as the angle between the directions of $\omega$ and $\tau$
as seen in the $\omega \pi$ rest frame, and 
\begin{align} 
\label{eq:A} 
A(s) =\,& \frac{G_F^2 S_{EW} |V_{ud}|^2}{2048 \pi^3 s^2 m_\tau^3} (s-m_\tau^2)^2 \lambda^{\frac{3}{2}}(s, M_\omega^2, M_\pi^2) \notag \\[1mm]
     & \times \Big\lbrace  \left( m_\tau^2 +s \right) |F_{V}(s)|^2 + 16 m_\tau \hat{\epsilon}_T \mathrm{Re} \left[ F_{V}(s) \left( F_{T3}^\ast(s)-F_{T2}^\ast(s) \right) \right] \Big\rbrace\,,\\[2mm]
\label{eq:B} 
B(s) =\,& -\frac{G_F^2 S_{EW} |V_{ud}|^2 }{128 \pi^3 s^2 m_\tau^2} (s-m_\tau^2)^2 \lambda(s, M_\omega^2, M_\pi^2) \notag \\[1mm]
     & \times \hat{\epsilon}_T \Big\lbrace \left( s-\Delta_{\omega \pi} \right) \mathrm{Re}\left[ F_{V}(s) F_{T2}^\ast(s)\right] + \left( s+\Delta_{\omega \pi} \right) \mathrm{Re}\left[ F_{V}(s) F_{T3}^\ast(s)\right] \Big\rbrace\,,  \\[2mm]  
\label{eq:C} 
C(s) =\,& \frac{G_F^2 S_{EW} |V_{ud}|^2}{2048 \pi^3 s^2 m_\tau^3} (s-m_\tau^2)^3 \lambda^{\frac{3}{2}}(s, M_\omega^2, M_\pi^2) |F_{V}(s)|^2\,.   
\end{align}
Here $\lambda(a,b,c)=a^2+b^2+c^2-2ab-2ac-2bc$ is the usual K{\"a}ll{\'e}n function, and the abbreviation $\Delta_{\omega \pi}=M_\omega^2 -M_\pi^2$ has been introduced in Eq.~\eqref{eq:B}. Since the NP contribution is assumed to be smaller than that of the SM, we have neglected all the quadratic terms in the tensor coefficient $\hat{\epsilon}_T$ in deriving Eqs.~\eqref{eq:A}--\eqref{eq:C}. In this approximation, the form factor $F_{T1}(s)$ does not appear at all, because it is always associated with the NP quadratic terms. When working at the linear order in $\epsilon_i$, the decay is also insensitive to the non-standard vector charged-current interaction, because the overall dependence on the combination $\epsilon_L+\epsilon_R$ cannot be isolated and is therefore generally subsumed in the determination of the Fermi constant $G_F^\mathrm{exp}$~\cite{Cirigliano:2009wk,Bhattacharya:2011qm,Miranda:2018cpf}. Throughout this paper, we shall use the abbreviation $G_F= G_F^0(1+\epsilon_L+\epsilon_R)$, with the superscript `exp' omitted.

Integrating further Eq.~\eqref{eq:dddw} over $\cos\theta$, one arrives at the differential decay width
\begin{align}\label{eq:dGamma}
	\frac{d\Gamma(\tau^- \to \omega \pi^- \nu_\tau)}{ds} =\,& \frac{G_F^{2} |V_{ud}|^2 (m_\tau^2-s)^2 S_{EW}}{1536 \pi^3 s^2 m_\tau^3} \lambda^{3/2}(s,M_\omega^2,M_\pi^2) \nonumber\\[0.1cm]
	&\hspace{-0.5cm} \times \Big\lbrace   (m_\tau^2 + 2s) |F_{V}(s)|^2 + 24 m_\tau \hat{\epsilon}_T \mathrm{Re}\left[ F_{V}(s)\left( F_{T3}^\ast(s) - F_{T2}^\ast(s) \right) \right]  \Big\rbrace\,,
\end{align}
and hence the spectral function~\cite{ARGUS:1986bzw,Tsai:1971vv,LopezCastro:1996xh} 
\begin{align}\label{eq:spec}
v(s) &= \frac{32 \pi^2 m_\tau^3}{G_F^{2} |V_{ud}|^2 (m_\tau^2 - s)^2 (m_\tau^2 + 2s)} \frac{d\Gamma(\tau^- \to \omega \pi^- \nu_\tau)}{ds} \nonumber\\[0.1cm]
     &=\frac{S_{EW}}{48 \pi s^2} \lambda^{\frac{3}{2}}(s,M_\omega^2,M_\pi^2) \left\{ |F_{V}(s)|^2 + \frac{24 m_\tau}{m_\tau^2 + 2s} \hat{\epsilon}_T \mathrm{Re}\left[ F_{V}(s)\left( F_{T3}^\ast(s) - F_{T2}^\ast(s) \right) \right] \right\}\,.
\end{align}
The branching ratio of $\tau^- \to \omega \pi^- \nu_\tau$ decay can then be obtained by integrating numerically Eq.~\eqref{eq:dGamma} over the variable $s$ from the threshold $(M_\omega+M_\pi)^2$ up to $m_\tau^2$, and multiplying the resulting expression with the $\tau$-lepton lifetime, 
\begin{align}
\mathcal{B}(\tau^- \to \omega \pi^- \nu_\tau)=\tau_\tau\int_{(M_\omega+M_\pi)^2}^{m_\tau^2}\frac{d\Gamma(\tau^- \to \omega \pi^- \nu_\tau)}{ds}\,ds\,,
\end{align}
with $\tau_\tau=290.3\times10^{-15}s$~\cite{ParticleDataGroup:2022pth}. The branching ratio and the spectral function $v(s)$ defined by Eq.~\eqref{eq:spec} are the only available observables that can be used to determine the R$\chi$T parameters and constrain the NP coefficient $\hat{\epsilon}_T$. In addition, one can study the NP effect in the angular distributions of the $\tau^- \to \omega \pi^- \nu_\tau$ decay, one of which is the so-called forward-backward asymmetry defined by~\cite{Chen:2021udz,Kimura:2012bwp,Beldjoudi:1994hi}
\begin{align}\label{eq:AFB}
	A_\mathrm{FB}(s) =\, &\frac{\int_{0}^{1}\frac{d^2\Gamma(\tau^- \to \omega \pi^- \nu_\tau)}{ds \, d\cos\theta}d\cos\theta - \int_{-1}^{0}\frac{d^2\Gamma(\tau^- \to \omega \pi^- \nu_\tau)}{ds \, d\cos\theta}d\cos\theta}{\int_{0}^{1}\frac{d^2\Gamma(\tau^- \to \omega \pi^- \nu_\tau)}{ds \, d\cos\theta}d\cos\theta + \int_{-1}^{0}\frac{d^2\Gamma(\tau^- \to \omega \pi^- \nu_\tau)}{ds \, d\cos\theta}d\cos\theta}\,.
\end{align}
Plugging Eqs.~\eqref{eq:dddw}--\eqref{eq:C} into Eq.~\eqref{eq:AFB} yields
\begin{align}
A_\mathrm{FB}(s) =\,& \frac{3B(s)}{6A(s)+2C(s)}  \notag\\[0.1cm]
	=\,& \frac{-12 m_\tau \hat{\epsilon}_T \Big\lbrace  \mathrm{Re}\left[ F_{V}(s)F_{T2}^\ast(s) \right] \left( s-\Delta_{\omega \pi} \right) + \mathrm{Re}\left[ F_{V}(s)F_{T3}^\ast(s) \right] \left( s+\Delta_{\omega \pi} \right) \Big\rbrace }{\lambda^{\frac{1}{2}}(s, M_\omega^2, M_\pi^2) \Big\lbrace  \left( 2s+m_\tau^2 \right) |F_{V}(s)|^2 +24 m_\tau \hat{\epsilon}_T \mathrm{Re}\left[ F_{V}(s)(F_{T3}^\ast(s)-F_{T2}^\ast(s)) \right] \Big\rbrace } \,.
\end{align}
Being proportional to the NP coefficient $\hat{\epsilon}_T$, the observable $A_\mathrm{FB}(s)$ is exactly zero in the SM and is therefore ideal to probe the existence of non-standard tensor interaction. Detailed numerical analysis by using the aforementioned observables to study the NP effect on the $\tau^-\to \omega\pi^-\nu_\tau$ decay will be performed in section~\ref{sec:numerical}.

\section{\boldmath Calculating the $\omega\pi$ form factors in R$\chi$T}
\label{sec:FFs}

In this section, we proceed to calculate the $\omega\pi$ form factors introduced in Eqs.~\eqref{eq:Had1} and \eqref{eq:Had2} in the context of R$\chi$T~\cite{Ecker:1988te,Ecker:1989yg,Ruiz-Femenia:2003jdx,Cirigliano:2004ue}, which extends the $\chi$PT~\cite{Weinberg:1978kz,Gasser:1983yg,Gasser:1984gg} by adding resonances as explicit degrees of freedom. As the $\omega\pi$ vector form factor has already been calculated in the R$\chi$T framework in Ref.~\cite{Guo:2008sh}, our main task in this section is to compute the tensor form factors in the same formalism.

\subsection[Relevant R$\chi$T Lagrangian]{\boldmath Relevant R$\chi$T Lagrangian}
\label{sec:L}

To facilitate the calculations of the $\omega\pi$ form factors in the R$\chi$T framework, let us firstly introduce all the necessary ingredients for constructing the relevant R$\chi$T Lagrangian, with the vector resonances described by the anti-symmetric tensor fields. Within the $\chi$PT and R$\chi$T frameworks, we need attach the external currents to the massless QCD Lagrangian $\mathcal{L}_\mathrm{QCD}^0$, allowing therefore to determine the relevant QCD currents:
\begin{align}
    \mathcal{L}_\mathrm{QCD} =\,& \mathcal{L}_\mathrm{QCD}^0 + \bar{\psi} \gamma_\mu (v^\mu + a^\mu \gamma_5)\psi - \bar{\psi} (s - i p \gamma_5)\psi + \bar{\psi} \sigma_{\mu\nu} \bar{t}^{\mu\nu}\psi\,,
\end{align}
where the external fields $v^\mu$, $a^\mu$, $s$, $p$, and $\bar{t}_{\mu\nu}$ are all $3\times 3$ Hermitian matrices in flavor space, with $v^\mu$ and $a^\mu$ chosen to be traceless in flavor space, whereas all the rest having in general a non-vanishing trace. Especially, the external tensor source $\bar{t}_{\mu\nu}$ includes both the octet and singlet currents~\cite{Cata:2007ns},
\begin{align}
    \bar{t}^{\mu\nu} = \sum_{a=0}^{8} \frac{\lambda^a}{2} \bar{t}^{{\mu\nu},a}\,,
\end{align}
where $\lambda^0=\sqrt{2/3}\,\mathbf{1}_{3\times3}$, and $\lambda^a$ ($a=1,\cdots,8$) are the Gell-Mann matrices with the convention $\text{Tr}(\lambda^a\lambda^b)=2\delta_{ab}$. These external fields allow us to compute the effective realization of the general Green functions of quark currents in a very straightforward way. Explicitly, once the relevant Lagrangian is fixed, the quark bilinear hadronization is then determined by taking the functional derivatives of the $\chi$PT and R$\chi$T actions with respect to the external fields and, afterwards, by setting all the external currents to zero.

The most relevant effective chiral Lagrangian at the leading order in the large $N_C$ expansion, with $N_C$ being the number of colors, is given by\footnote{When writing Eq.~\eqref{eq:RchiT} without the $\mathcal{O}(p^4)$ and higher-order chiral operators with Goldstone bosons only, we have assumed the resonance saturation of the couplings associated with these operators.}
\begin{align}\label{eq:RchiT}
	\mathcal{L}=\mathcal{L}_\chi^{(2)}  + \mathcal{L}_{kin}(V) + \mathcal{L}_{2V} + \mathcal{L}_{VVP}+ \mathcal{L}_{VJP}\,.
\end{align}
The first term of Eq.~\eqref{eq:RchiT} is the leading order ($\mathcal{O}(p^2)$) Lagrangian of $\chi$PT given by~\cite{Gasser:1983yg,Gasser:1984gg}
\begin{align}\label{eq:chiPT}
\mathcal{L}_\chi^{(2)}=\frac{F^2}{4}\left\langle  u_\mu u^\mu + \chi_+ \right\rangle\,,
\end{align}
where $F$ is the decay constant of the charged pion, and $\langle\cdots\rangle$ represents the trace in flavor space. Here $u_\mu =i \left[u^\dagger(\partial_\mu-i r_\mu)u-u(\partial_\mu-i l_\mu)u^\dagger\right]=i u^\dagger D_\mu U u^\dagger$ and $\chi_{\pm}=u^\dagger \chi u^\dagger \pm u \chi^\dagger u$, where $u=\sqrt{U}=e^{i\frac{\Phi}{\sqrt{2}F}}$ is a unitary matrix in flavor space and denotes a non-linear representation of the lowest-lying pseudo-scalar nonet with
\begin{align}
\Phi= \begin{pmatrix}
		\frac{\pi^0}{\sqrt{2}}+\frac{1}{\sqrt{6}}\eta_8+\frac{1}{\sqrt{3}}\eta_0 & \pi^+ & K^+   \\[0.2cm]
		\pi^- & -\frac{\pi^0}{\sqrt{2}} +\frac{1}{\sqrt{6}}\eta_8+\frac{1}{\sqrt{3}}\eta_0 & K^0 \\[0.2cm]
		K^- & \bar{K}^0 & -\frac{2}{\sqrt{6}}\eta_8 +\frac{1}{\sqrt{3}}\eta_0
	\end{pmatrix} \,,
\end{align}
and $\chi=2B_0 (s+ip)$, with $B_0$ being a constant related to the quark condensate. The covariant derivative for the pseudo-scalar nonet is defined by
\begin{align}
    D_\mu U= \partial_\mu U - i r_\mu U + i U l_\mu\,, \qquad 
    D_\mu U^\dagger= \partial_\mu U^\dagger + i U^\dagger r_\mu - i l_\mu  U^\dagger\,,
\end{align}
from which the field strength tensors arise naturally for the right- and left-handed fields,
\begin{align}
    \left[D^\mu, D^\nu\right] X = i X F_L^{\mu \nu} - i F_R^{\mu \nu} X\,,
\end{align}
with $F_L^{\mu\nu} = \partial^\mu l^\nu-\partial^\nu l^\mu-i\left[l^\mu, l^\nu\right]$, and $F_R^{\mu\nu} = \partial^\mu r^\nu-\partial^\nu r^\mu-i\left[r^\mu, r^\nu\right]$. The external Hermitian matrix fields $r_\mu=v_\mu+a_\mu$, $l_\mu=v_\mu-a_\mu$, $s$ and $p$ promote the global $SU(3)_L\otimes SU(3)_R$ chiral symmetry of the Lagrangian to a local one, which then breaks spontaneously down to the diagonal subgroup $SU(3)_V$. 

The second and the third term of Eq.~\eqref{eq:RchiT} describe the kinetics and the interaction of the vector resonances with the Goldstone fields, which are given, respectively, by~\cite{Ecker:1988te,Ecker:1989yg}\footnote{It should be noted that the traces over pseudo-scalar and vector nonets in Eqs.~\eqref{eq:chiPT}, \eqref{eq:kin}, \eqref{eq:2V}, \eqref{eq:VVP} and \eqref{eq:VJP} are non-zero, which implies that there are some extra operators in terms of the traces over these nonets like $\langle V_{\mu\nu}\rangle\langle V^{\mu\nu}\rangle\langle \cdots \rangle$. Nevertheless, as they are irrelevant to the $\tau^- \to \omega\pi^-\nu_\tau$ decay, we have simply discarded these terms in the above expressions.}
\begin{align}
	\mathcal{L}_{kin}(V)=&-\frac{1}{2} \langle \nabla^\lambda V_{\lambda \mu} \nabla_\nu V^{\nu \mu} - \frac{M_V^2}{2}V^{\mu\nu} V_{\mu\nu}  \rangle\,,\label{eq:kin}\\[0.2cm]
	\mathcal{L}_{2V} =&\, \frac{F_V}{2\sqrt{2}}\left\langle  V_{\mu\nu} f_+^{\mu\nu} \right\rangle  + \frac{i G_V}{2\sqrt{2}}\left\langle  V_{\mu\nu} [u^\mu, u^\nu] \right\rangle\,,\label{eq:2V}
\end{align}
where $f_{\pm}^{\mu\nu}=uF_L^{\mu\nu}u^\dagger \pm u^\dagger F_R^{\mu\nu}u$, and the vector nonet formulated in the anti-symmetric tensor form is given by
\begin{align}\label{eq:Vmn}
V_{\mu\nu}=\begin{pmatrix}
		\frac{\rho^0}{\sqrt{2}}+\frac{\omega}{\sqrt{2}} & \rho^+ & K^{\ast +}   \\[0.2cm]
		\rho^- & -\frac{\rho^0}{\sqrt{2}}+\frac{\omega}{\sqrt{2}} & K^{\ast 0} \\[0.2cm]
		K^{\ast -} & \bar{K}^{\ast 0} & \phi 
	\end{pmatrix}_{\mu\nu}\,.
\end{align}
Here $M_V$ is the mass of the $\rho$ meson, and $F_V$ and $G_V$ the corresponding resonance couplings that can be determined from the $\rho$ decay rates~\cite{Ecker:1988te}. The covariant derivative $\nabla_\mu V=\partial_\mu V+[\Gamma_\mu,V]$, with the chiral connection $\Gamma_\mu=\frac{1}{2}\left\{u^\dagger(\partial_\mu -ir_\mu)u + u(\partial_\mu - i l_\mu)u^\dagger\right\}$, is defined in such a way that it also transforms as an octet under the action of the chiral group. We have also assumed an ideal mixing for the vector resonances $\omega$ and $\phi$, which means that
\begin{align}
\omega_1=\sqrt{\frac{2}{3}}\omega-\sqrt{\frac{1}{3}}\phi\,,\qquad
\omega_8=\sqrt{\frac{1}{3}}\omega+\sqrt{\frac{2}{3}}\phi\,.
\end{align}

The remaining terms of Eq.~\eqref{eq:RchiT} denote the odd-intrinsic-parity sector involving one pseudo-scalar and one vector resonance. Explicitly, we have~\cite{Ruiz-Femenia:2003jdx}\footnote{Due to the $G$-parity conservation, the Lagrangian $\mathcal{L}_{VAP}$, which describes the interactions among the pseudo-scalar, vector, and axial-vector resonances~\cite{Cirigliano:2004ue}, has no contribution to the $\tau^- \to \omega\pi^-\nu_\tau$ decay, and hence is not given here. Nevertheless, it will contribute to other $\tau^- \to (VP)^-\nu_\tau$ decays~\cite{Guo:2008sh}.}
\begin{align}
	\mathcal{L}_{VVP} =\,&\, d_1 \epsilon_{\mu\nu\rho\sigma} \left\langle  \left\lbrace V^{\mu\nu}, V^{\rho\alpha}\right\rbrace  \nabla_\alpha u^\sigma \right\rangle  + id_2 \epsilon_{\mu\nu\rho\sigma} \left\langle  \left\lbrace V^{\mu\nu},V^{\rho\sigma}\right\rbrace  \chi_- \right\rangle  \nonumber \\[0.1cm]	
	&+ d_3 \epsilon_{\mu\nu\rho\sigma} \left\langle  \left\lbrace \nabla_\alpha V^{\mu\nu},V^{\rho\alpha}\right\rbrace  u^\sigma \right\rangle  + d_4 \epsilon_{\mu\nu\rho\sigma} \left\langle  \left\lbrace \nabla^\sigma V^{\mu\nu},V^{\rho\alpha}\right\rbrace  u_\alpha \right\rangle \,, \label{eq:VVP}\\[2mm]
	\mathcal{L}_{VJP} =\,& \frac{c_1}{M_V}\epsilon_{\mu\nu\rho\sigma} \left\langle  \left\lbrace V^{\mu\nu},f_+^{\rho\alpha}\right\rbrace  \nabla_\alpha u^\sigma \right\rangle  + \frac{c_2}{M_V}\epsilon_{\mu\nu\rho\sigma} \left\langle  \left\lbrace V^{\mu\alpha},f_+^{\rho\sigma}\right\rbrace  \nabla_\alpha u^\nu \right\rangle  \nonumber \\[0.1cm]
	&+ \frac{ic_3}{M_V}\epsilon_{\mu\nu\rho\sigma} \left\langle  \left\lbrace V^{\mu\nu},f_+^{\rho\sigma}\right\rbrace  \chi_- \right\rangle  + \frac{ic_4}{M_V}\epsilon_{\mu\nu\rho\sigma} \left\langle  V^{\mu\nu} \left[ f_-^{\rho\sigma}, \chi_+\right] \right\rangle  \nonumber \\[0.1cm]
	&+ \frac{c_5}{M_V}\epsilon_{\mu\nu\rho\sigma} \left\langle  \left\lbrace \nabla_\alpha V^{\mu\nu},f_+^{\rho\alpha}\right\rbrace  u^\sigma \right\rangle  + \frac{c_6}{M_V}\epsilon_{\mu\nu\rho\sigma} \left\langle  \left\lbrace \nabla_\alpha V^{\mu\alpha},f_+^{\rho\sigma}\right\rbrace  u^\nu \right\rangle  \nonumber \\[0.1cm]
	&+ \frac{c_7}{M_V}\epsilon_{\mu\nu\rho\sigma} \left\langle  \left\lbrace \nabla^\sigma V^{\mu\nu},f_+^{\rho\alpha}\right\rbrace  u_\alpha \right\rangle\, , \label{eq:VJP}
\end{align}
where $d_i$ and $c_i$ are the relevant resonance couplings. If a second heavier nonet of vector resonances, $V_1$, is involved in the decay, one should include, besides a replica of the Lagrangians specified by Eqs.~\eqref{eq:kin}, \eqref{eq:2V}, \eqref{eq:VVP} and \eqref{eq:VJP}, the following new Lagrangian~\cite{Mateu:2007tr}:
\begin{align} \label{eq:LVV1P}
	\mathcal{L}_{VV_1P} =\,& d_a \epsilon_{\mu\nu\rho\sigma} \left\langle  \left\lbrace V^{\mu\nu}, V_1^{\rho\alpha}\right\rbrace  \nabla_\alpha u^\sigma \right\rangle  + d_b \epsilon_{\mu\nu\rho\sigma} \left\langle  \left\lbrace V^{\mu\alpha},V_1^{\rho\sigma}\right\rbrace  \nabla_\alpha u^\nu \right\rangle  \nonumber\\[0.1cm]
	&+ d_c \epsilon_{\mu\nu\rho\sigma} \left\langle  \left\lbrace \nabla_\alpha V^{\mu\nu},V_1^{\rho\alpha}\right\rbrace  u^\sigma \right\rangle  + d_d \epsilon_{\mu\nu\rho\sigma} \left\langle  \left\lbrace \nabla_\alpha V^{\mu\alpha},V_1^{\rho\sigma}\right\rbrace  u^\nu \right\rangle  \nonumber \\[0.1cm]
	&+ d_e \epsilon_{\mu\nu\rho\sigma} \left\langle  \left\lbrace \nabla^\sigma V^{\mu\nu},V_1^{\rho\alpha}\right\rbrace  u_\alpha \right\rangle  + id_f \epsilon_{\mu\nu\rho\sigma} \left\langle  \left\lbrace V^{\mu\nu}, V_1^{\rho\sigma}\right\rbrace  \chi_- \right\rangle\,,
\end{align}
where $V_1^{\mu\nu}$ denotes the second nonet of vector resonances and has a similar form as of $V^{\mu\nu}$ defined by Eq.~\eqref{eq:Vmn}, up to a different mass. Note that the first five terms of Eq.~\eqref{eq:LVV1P} describe the interactions of the two vector resonances from different multiplets with the lowest-lying pseudo-scalars, while the last term contains the vertices with these two vector nonets and one external pseudo-scalar source~\cite{Mateu:2007tr}. 

As the R$\chi$T Lagrangian presented above does not involve the external tensor sources, we are still unable to calculate the $\omega \pi$ tensor form factors with just these operators. In the presence of a non-standard tensor interaction, We must add to Eq.~\eqref{eq:RchiT} the following terms involving the tensor building blocks $t^{\mu\nu}_{\pm}$:
\begin{align}
	\mathcal{L}_{VT} =\,& F^T_V \left\langle  V_{\mu\nu} t^{\mu\nu}_+ \right\rangle\,, \label{eq:VT} \\[0.2cm]
	\mathcal{L}_{V T P} =\, & b_1 \epsilon_{\mu \nu \rho \sigma}\left\langle\left\{V^{\mu \nu}, t_{+}^{\rho \alpha}\right\} \nabla_\alpha u^\sigma\right\rangle +b_2 \epsilon_{\mu \nu \rho \sigma}\langle \{V^{\mu \alpha},t_{+}^{\nu\rho}\} \nabla_\alpha u^\sigma\rangle  \nonumber \\[1mm]
	&+i b_3 \epsilon_{\mu \nu \rho \sigma}\left\langle\left\{V^{\mu \nu}, t_{+}^{\rho \sigma}\right\} \chi_{-}\right\rangle +i b_4 \epsilon_{\mu \nu \rho \sigma}\left\langle\left\{V^{\mu \nu}, t_{-}^{\rho \sigma}\right\} \chi_{+}\right\rangle \nonumber \\[1mm]
	& +b_{5} \epsilon_{\mu \nu \rho \sigma}\left\langle\left\{\nabla_\alpha V^{\mu \nu}, t_{+}^{\rho \alpha}\right\} u^\sigma\right\rangle +b_{6} \epsilon_{\mu \nu \rho \sigma}\left\langle\left\{\nabla_\alpha V^{\mu \alpha}, t_{+}^{\nu\rho}\right\} u^\sigma\right\rangle \nonumber \\[1mm]
	&+b_{7} \epsilon_{\mu \nu \rho \sigma}\left\langle\left\{\nabla^\mu V^{\nu \rho}, t_{+}^{\sigma \alpha}\right\} u_\alpha\right\rangle +i b_8 \epsilon_{\mu \nu \rho \sigma}\left\langle V^{\mu \nu}t_{-}^{\rho \sigma}\right\rangle\langle \chi_{+} \rangle \nonumber \\[1mm]
	&+b_9\epsilon_{\mu \nu \rho \sigma}\langle V^{\mu \nu} \nabla_\alpha u^\rho\rangle\langle t_{+}^{\sigma \alpha} \rangle +b_{10} \epsilon_{\mu \nu \rho \sigma}\langle V^{\mu\nu}\nabla^{\rho}u_\alpha \rangle\langle  t_{+}^{\sigma\alpha} \rangle \nonumber \\[1mm]
	&+i b_{11} \epsilon_{\mu \nu \rho \sigma}\left\langle V^{\mu \nu}  \chi_{-}\right\rangle\langle t_{+}^{\rho \sigma}\rangle +i b_{12} \epsilon_{\mu \nu \rho \sigma}\left\langle V^{\mu \nu}\chi_{+}\right\rangle\langle t_{-}^{\rho \sigma} \rangle \nonumber \\[1mm]
	& + b_{13} \epsilon_{\mu \nu \rho \sigma}\left\langle\nabla_\alpha V^{\mu \nu} u^\rho\right\rangle\langle t_{+}^{\sigma \alpha} \rangle +b_{14} \epsilon_{\mu \nu \rho \sigma}\left\langle\nabla_\alpha V^{\mu \alpha}u^\nu\right\rangle\langle t_{+}^{\rho \sigma}\rangle \nonumber\\[1mm]
	& +b_{15} \epsilon_{\mu \nu \rho \sigma}g_{\alpha\beta}\left\langle\nabla^\mu V^{ \nu\alpha} u^\rho\right\rangle\langle t_{+}^{\sigma \beta} \rangle\,, \label{eq:VTP}
\end{align}
where $F_V^T$ and $b_i$ ($i=1,\cdots,15$) represent the corresponding resonance couplings, and $t_{\pm}^{\mu\nu} = u^\dagger t^{\mu\nu} u^\dagger \pm u {t^{\mu\nu}}^\dagger u$, with $t^{\mu\nu}$ and $t^{\mu\nu\dagger}$ being the chiral external tensor sources. They are related to the original external tensor field $\bar{t}^{\mu\nu}$ via~\cite{Cata:2007ns}
\begin{align}
 \bar{t}^{\mu\nu}=P_L^{\mu\nu\lambda\rho}t_{\lambda\rho}+P_R^{\mu\nu\lambda\rho}t_{\lambda\rho}^\dagger\,, \qquad
 t^{\mu\nu(\dagger)}=P_{L(R)}^{\mu\nu\lambda\rho} \bar{t}_{\lambda\rho}\,,
\end{align}
in which the projection operators $P_R^{\mu\nu\lambda\rho}=\frac{1}{4} (g^{\mu\lambda} g^{\nu\rho}-g^{\nu\lambda} g^{\mu\rho}-i\epsilon^{\mu\nu\lambda\rho})$ and $ P_L^{\mu\nu\lambda\rho}=(P_R^{\mu\nu\lambda\rho})^\dagger$ have been introduced. Likewise, if there involves a second nonet of vector resonances, $V_1$, in the decay, one should then add two similar Lagrangians as given by Eqs.~\eqref{eq:VT} and \eqref{eq:VTP} to describe its contribution, by simply replacing $V$ with $V_1$. While the Lagrangian $\mathcal{L}_{VT}$ has been written down explicitly in Refs.~\cite{Mateu:2007tr,Husek:2020fru,Chen:2019vbr}, $\mathcal{L}_{VTP}$ is constructed for the first time in this work. For more details about the construction of the R$\chi$T Lagrangian with these external tensor sources, we refer the readers to appendix~\ref{app:constructVTP}.

\subsection[The $\omega\pi$ vector and tensor form factors]{\boldmath The $\omega\pi$ vector and tensor form factors}
\label{subsec:ff}

With the relevant R$\chi$T Lagrangian constructed above, a direct calculation yields the explicit expressions of the $\omega\pi$ form factors $F_{V}(s)$,\footnote{Note that the same expression of the vector form factor $F_V(s)$ as given by Eq.~\eqref{eq:FV} has already been presented in Ref.~\cite{Guo:2008sh}.} $F_{T1}(s)$, $F_{T2}(s)$, and $F_{T3}(s)$, with
\begin{align}
	F_{V}(s)=& -\frac{4}{F M_V M_\omega} \left[(c_1+c_2+8c_3-c_5)M_\pi^2 +(c_2-c_1+c_5-2c_6)M_\omega^2+(c_1-c_2+c_5)s \right] \nonumber \\[1mm]
	&+\frac{4\sqrt{2} F_V}{F M_\omega} \left[ d_{12} M_\pi^2 + d_3 (s+\Delta_{\omega\pi}) \right] D_\rho(s)\nonumber \\[1mm]
	&+\frac{2\sqrt{2} F_{V_1}}{F M_\omega} \left[d_m M_\pi^2 + d_M M_\omega^2+d_s s\right] D_{\rho^\prime}(s)\,, \label{eq:FV}\\[2mm]
	F_{T1}(s)=&-\frac{2 (b_1-b_2)}{F M_V M_\omega} +\frac{4 F^T_V}{F M_\omega M_\rho^2} \left[ d_{12} M_\pi^2 + d_3 \left( \Delta_{\omega\pi} +M_\rho^2 \right) +d_4 \left( M_\rho^2 -s \right) \right] D_\rho(s)  \nonumber \\[1mm]
	&+\frac{2 F^T_{V_1}}{F M_\omega M_{\rho^\prime}^2} \left[ d_M M_\omega^2 +d_m M_\pi^2 -\left( d_a-d_b-d_c \right) s + 2 \left( d_a-d_b \right)M_{\rho^\prime}^2 \right] D_{\rho^\prime}(s)\,,  \label{eq:FT1}\\[2mm]
	F_{T2}(s)=& \frac{1}{F M_V M_\omega} \left[\left( b_1-b_2 \right) (s-\Sigma_{\omega\pi}) +2 \left( b_5-b_6 \right) M_\omega^2 \right]  \nonumber\\[1mm]
	& -\frac{2 F^T_V}{F M_\omega M_\rho^2}\left[ d_{12} M_\pi^2 (s+\Delta_{\omega\pi}) -d_3\left( s\Sigma_{\omega\pi}-\Delta_{\omega\pi}^2 + M_\rho^2 \left( 2 M_\omega^2 +s+\Delta_{\omega\pi} \right)  \right) \right. \nonumber\\[1mm]
	&\left. - d_4 \left( s^2 -s\left( \Sigma_{\omega\pi} +M_\rho^2 \right) +M_\rho^2 \Sigma_{\omega\pi} \right)  \right] D_\rho(s) \nonumber\\[1mm]
	&+\frac{F^T_{V_1}}{F M_\omega M_{\rho^\prime}^2}\left[ (s-\Sigma_{\omega\pi}) \left( d_M M_\omega^2 -d_m M_\pi^2 -\left( d_a-d_b-d_c \right)s -2\left( d_a-d_b \right)M_{\rho^\prime}^2 \right)  \right. \nonumber\\[1mm]
	&\left. - 4M_\omega^2 \left( \left( d_a-d_c+d_d+4 d_f \right) M_\pi^2 +\left( d_c-d_d \right)M_{\rho^\prime}^2 \right)  \right] D_{\rho^\prime}(s)\,, \label{eq:FT2}\\[2mm]
	F_{T3}(s)=& -\frac{1}{F M_V M_\omega} \left[ b_5 (s-\Sigma_{\omega\pi}) +2 \left( b_1+4b_3+4b_4+4b_8 \right) M_\pi^2 \right]  \nonumber\\[1mm]
	& -\frac{2 F^T_V}{F M_\omega M_\rho^2}\left[ d_{12} M_\pi^2 \left( (s+\Delta_{\omega\pi}) -2 M_\rho^2 \right) - d_4 \left( s^2 +s\left( \Sigma_{\omega\pi} +M_\rho^2 \right) -M_\rho^2 \Sigma_{\omega\pi}  \right)  \right. \nonumber\\[1mm]
	&\left. -d_3\left( s\Sigma_{\omega\pi}-\Delta_{\omega\pi}^2 + M_\rho^2 \left( 2 M_\omega^2 +s+\Delta_{\omega\pi} \right)  \right)  \right] D_\rho(s) \nonumber\\[1mm]
	&+\frac{F^T_{V_1}}{F M_\omega M_{\rho^\prime}^2}\left[ (s-\Sigma_{\omega\pi}) \left( d_M M_\omega^2 -d_m M_\pi^2 -\left( d_a-d_b-d_c \right)s +2 d_c M_{\rho^\prime}^2 \right)  \right. \nonumber\\[1mm]
	&\left. - 4M_\pi^2 \left( \left( d_a-d_c+d_d+4 d_f \right) M_\omega^2 -\left( d_a+4 d_f \right)M_{\rho^\prime}^2 \right)  \right] D_{\rho^\prime}(s)\,, \label{eq:FT3}
\end{align}
where $D_V(s)$ denotes the propagator of the intermediate vector resonance $V$ with an energy-dependent decay width $\Gamma_V(s)$, whose explicit expression can be found in Refs.~\cite{Guo:2008sh,Jamin:2006tk}. For convenience, we have also introduced the combinations,
\begin{align} \label{eq:shorthand}
	d_m=& d_a+d_b-d_c+8 d_f\,,  \\[1mm]
	d_M=& d_b-d_a+d_c-2 d_d\,, \\[1mm]
	d_s=& d_a-d_b+d_c\,,  \\[1mm]
	d_{12}=& d_1+8 d_2\,,  \\[1mm]
	\Sigma_{\omega\pi}=& M_\omega^2 +M_\pi^2\,.
\end{align}

It is apparent that too many undetermined resonance couplings are involved in these form factors, rendering therefore the theory to be less predictive. To exploit fully these form factors in our numerical analysis, we shall firstly try to remove all the unnecessary and redundant parameters. As the R$\chi$T itself cannot tell us anything about these couplings, we shall resort to other available constraints to pin down their values. To this end, we can analyze the QCD-ruled short-distance properties of various Green functions and then compare them with the same objects built from the effective action with explicit resonance degrees of freedom, to impose restrictions on these unknown couplings~\cite{Knecht:2001xc,Bijnens:2003rc,Ruiz-Femenia:2003jdx,Mateu:2007tr,Roig:2013baa}. Especially, the QCD short-distance behavior demands that the form factors should vanish smoothly for $s\to\infty$, which, although being not derived from first principles, is heuristically inferred and phenomenologically supported~\cite{Brodsky:1973kr,Lepage:1979zb,Lepage:1980fj}. Following this procedure, we obtain the following constraints on the resonance couplings: 
\begin{align}
	\label{eq:c125}  c_1-c_2+c_5=&c_1+4 c_3=d_4=d_a-d_b-d_c=b_1-b_2=b_5= 0\,,  \\[1mm]
	\label{eq:c65}  c_6-c_5=& \frac{2 d_3 F_V+d_s F_{V_1}}{2 \sqrt{2}M_\omega^2}M_V\,,\\[1mm]
	\label{eq:b6}  b_6=& \frac{F_{V}^T }{M_\omega^2 M_{\rho}^2} \left[ d_{12} M_\pi^2 -d_3 \left( \Sigma_{\omega\pi} -M_\rho^2 \right) \right]  \nonumber \\[1mm]
	&+\frac{F_{V_1}^T}{M_\omega^2 M_{\rho^\prime}^2} \left[ d_d M_\omega^2 +\left( d_b +4 d_f \right) M_\pi^2 +\left( d_a-d_b \right) M_{\rho^\prime}^2 \right]\,,   \\[1mm]	
	\label{eq:b1348} b_1+4\left( b_3+b_4+b_8 \right)=& \frac{F_{V}^T }{M_\pi^2 M_{\rho}^2} \left[ d_{12} M_\pi^2 -d_3 \left( \Sigma_{\omega\pi} +M_\rho^2 \right) \right]  \nonumber \\[1mm]
    &+\frac{F_{V_1}^T}{M_\pi^2 M_{\rho^\prime}^2} \left[ d_d M_\omega^2 +\left( d_b +4 d_f \right) M_\pi^2 - d_c M_{\rho^\prime}^2 \right]\,.
\end{align}
After taking into account all these relations, we can rewrite the $\omega\pi$ form factors as
\begin{align}
	\label{eq:V(s)} F_{V}(s)=& \frac{2 \sqrt{2}}{F M_\omega} \left( 2 d_3 F_V +d_s F_{V_1} \right) +\frac{4\sqrt{2} F_V}{F M_\omega} \left[ d_{12} M_\pi^2 + d_3 (s+\Delta_{\omega\pi}) \right] D_\rho(s) \nonumber \\[1mm]
	&+\frac{2\sqrt{2} F_{V_1}}{F M_\omega} \left[d_m M_\pi^2 + d_M M_\omega^2+d_s s\right] D_{\rho^\prime}(s)\,, \\[2mm]
	\label{eq:T_1(s)} F_{T1}(s)=& \frac{4 F^T_V}{F M_\omega M_\rho^2} \left[ d_{12} M_\pi^2 + d_3 \left( \Delta_{\omega\pi} +M_\rho^2 \right) \right] D_\rho(s)  \nonumber \\[1mm]
	&-\frac{4 F^T_{V_1}}{F M_\omega M_{\rho^\prime}^2} \left[ d_d M_\omega^2 -\left( d_b +4 d_f \right) M_\pi^2 -\left( d_a-d_b \right) M_{\rho^\prime}^2 \right] D_{\rho^\prime}(s)\,,  \\[2mm]
	\label{eq:T_2(s)} F_{T2}(s)=& -\frac{2 F^T_V}{F M_\omega M_\rho^2}  \left\lbrace \left[ d_{12} M_\pi^2 -d_3 \left( \Sigma_{\omega\pi} -M_\rho^2 \right) \right] \right. \nonumber\\[1mm]
	&\left. + \left[ d_{12} M_\pi^2 (s+\Delta_{\omega\pi}) -d_3\left( s\Sigma_{\omega\pi}-\Delta_{\omega\pi}^2 + M_\rho^2 \left( 2 M_\omega^2 +s+\Delta_{\omega\pi} \right)  \right) \right] D_\rho(s) \right\rbrace \nonumber\\[1mm]	
	& -\frac{2 F^T_{V_1}}{F M_\omega M_{\rho^\prime}^2} \left\lbrace \left[ d_d M_\omega^2 +\left( d_b +4 d_f \right) M_\pi^2 +\left( d_a-d_b \right) M_{\rho^\prime}^2 \right] \left( 1+(s-\Sigma_{\omega\pi}) D_{\rho^\prime}(s) \right) \right. \nonumber \\[1mm]
	&\left. + 2 M_\omega^2 \left[ \left( d_b+d_d+4 d_f \right) M_\pi^2 +\left( d_a-d_b-d_d \right) M_{\rho^\prime}^2 \right] D_{\rho^\prime}(s) \right\rbrace\,, \\[2mm]
	\label{eq:T_3(s)} F_{T3}(s)=& -\frac{2 F^T_V}{F M_\omega M_\rho^2} \left\lbrace \left[ d_{12} M_\pi^2 -d_3 \left( \Sigma_{\omega\pi} +M_\rho^2 \right) \right] \right. \nonumber\\[1mm]
	&\left. +\left[ d_{12} M_\pi^2 \left( s+\Delta_{\omega\pi} -2 M_\rho^2 \right) -d_3 \left( s\Sigma_{\omega\pi}-\Delta_{\omega\pi}^2 + M_\rho^2(s-\Sigma_{\omega\pi}) \right) \right] D_{\rho}(s) \right\rbrace  \nonumber \\[1mm]
	&-\frac{2 F^T_{V_1}}{F M_\omega M_{\rho^\prime}^2} \left\lbrace \left[ d_d M_\omega^2 +\left( d_b +4 d_f \right) M_\pi^2 -\left( d_a-d_b \right) M_{\rho^\prime}^2 \right] \left( 1+(s-\Sigma_{\omega\pi}) D_{\rho^\prime}(s) \right) \right. \nonumber \\[1mm]
	&\left. + 2 M_\pi^2 \left[ \left( d_b+d_d+4 d_f \right) M_\omega^2 -\left( d_a+4 d_f \right) M_{\rho^\prime}^2 \right] D_{\rho^\prime}(s) \right\rbrace\,.
\end{align}
Here an interesting observation is that, since all of the (combinations of) couplings $b_i$ and $c_i$ can be expressed in terms of $d_i$, the $\omega\pi$ form factors can be uniquely determined once the couplings $d_i$ are known. 

\section{Numerical analysis and discussions}
\label{sec:numerical}

\subsection{Inputs and numerical results for the form factors}

\begin{table}[t]
	\tabcolsep 0.042in
	\renewcommand\arraystretch{1.5}
	\begin{center}
			\vspace{0.2cm}
			\begin{tabular}{|c|c|c|c|c|}
					\hline\hline
					$G_F~[\text{GeV}^{-2}]$~\cite{ParticleDataGroup:2022pth} & $\tau_\tau~[s]$~\cite{ParticleDataGroup:2022pth} & $V_{ud}$~\cite{ParticleDataGroup:2022pth} & $S_{EW}$~\cite{Erler:2002mv} & $F~[\text{MeV}]$~\cite{FlavourLatticeAveragingGroupFLAG:2021npn} \\
					\hline
					$1.1663788(6)\times10^{-5}$ & $290.3\times10^{-15}$ & $0.97373(31)$ & $1.0201(3)$ & $92.21(1)$  \\
					\hline
					$F_{V}~[\text{MeV}]$~\cite{Chen:2022gdi} & $G_{V}~[\text{MeV}]$~\cite{Chen:2022gdi} & $F_{V_1}~[\text{MeV}]$~\cite{Roig:2014uja} &   $F_{V}^T~[\text{GeV}^2]$~\cite{Husek:2020fru} & $F_{V_1}^T~[\text{GeV}^2]$~\cite{Cata:2008zc}   \\
					\hline
					 $153.27$ & $55.06$ & $-108.38$ &   $0.1147$ & $-0.05548$  \\
					\hline
					$m_{\tau}~[\text{MeV}]$~\cite{ParticleDataGroup:2022pth} & $M_{\omega}~[\text{MeV}]$~\cite{ParticleDataGroup:2022pth} & $M_{\pi^-}~[\text{MeV}]$~\cite{ParticleDataGroup:2022pth} & $M_{\pi^0}~[\text{MeV}]$~\cite{ParticleDataGroup:2022pth} & $M_{K^0}~[\text{MeV}]$~\cite{ParticleDataGroup:2022pth}    \\
					\hline
					$1776.86(12)$ & $782.66(13)$ & $139.57039(18)$ & $134.9768(5)$  & $ 497.611(13)$  \\
					\hline
					$M_\rho~[\text{MeV}]$~\cite{ParticleDataGroup:2022pth} & $M_{\rho^\prime}~[\text{MeV}]$~\cite{ParticleDataGroup:2022pth} & $\Gamma_{\rho^\prime}~[\text{MeV}]$~\cite{ParticleDataGroup:2022pth} & $d_{12}$~\cite{Ruiz-Femenia:2003jdx} &  $c_3$~\cite{Wang:2023njt}    \\
					\hline
					$775.26(23)$ & $1465(25)$ & $400(60)$ & $-0.03085$ & $0.00186$  \\
					\hline \hline
				\end{tabular}
				\caption{Summary of the input parameters used in our numerical analysis throughout this work. \label{tab:input}} 
		\end{center}
\end{table}

For convenience, we collect in Table~\ref{tab:input} the values of the input parameters used in our numerical analysis throughout this work. Let us now explain in detail how to determine the numerical results of the resonance couplings needed to obtain the $\omega\pi$ form factors given by Eqs.~\eqref{eq:V(s)}--\eqref{eq:T_3(s)}. Firstly, the values of the two couplings $F_V$ and $G_V$ are taken from the averages of the four fit results presented in Ref.~\cite{Chen:2022gdi}, and $F_{V_1}$ is estimated by using the relation $F_{V_n}=(-1)^n F_V/\sqrt{n+1}$~\cite{Roig:2014uja}. 

To estimate $G_{V_1}$, we follow a matching procedure from R$\chi$T to $\chi$PT by integrating out the resonance degrees of freedom, and express it in terms of the low-energy constants (LECs) of $\chi$PT~\cite{Ecker:1988te,Cirigliano:2006hb,Ruiz-Femenia:2003jdx,Kampf:2011ty}. Since the vector resonance exchange from two $\mathcal{O}(p^2)$ vertices described by $\mathcal{L}_{2V}$ will automatically produce a contribution of $\mathcal{O}(p^4)$, it should be matched to the $\mathcal{O}(p^4)$ low-energy $\chi$PT Lagrangian~\cite{Ecker:1988te}. Taking into account the contributions from the first two vector resonances, $V$ and $V_1$, one obtains the following relations~\cite{Ecker:1988te}:
\begin{align}
L_{1}^V=&\frac{G_{V}^{2}}{8M_{V}^{2}}+\frac{G_{V_1}^{2}}{8M_{V_1}^{2}}\,, &	L_{2}^V=&\, 2L_{1}^V\,,&  L_{3}^V=&-6L_{1}^V\,, \nonumber\\[0.1cm]
L_{9}^V=&\frac{F_{V}G_{V}}{2M_{V}^{2}}+\frac{F_{V_1}G_{V_1}}{2M_{V_1}^{2}}\,, & L_{10}^V=&-\frac{F_{V}^{2}}{4M_{V}^{2}}-\frac{F_{V_1}^{2}}{4M_{V_1}^{2}}\,, & H_{1}^V=&-\frac{F_{V}^2}{8M_{V}^{2}}-\frac{F_{V_1}^2}{8M_{V_1}^{2}}\,, 
\label{eq:lecr}
\end{align}
where $L_1^V$, $L_2^V$, $L_3^V$, $L_9^V$, $L_{10}^V$, and $H_1^V$ are the LECs of $\mathcal{O}(p^4)$ $\chi$PT Lagrangian~\cite{Ecker:1988te}, with the superscript $V$ labelling the contribution due to the exchange of spin-1 vector resonances. The LECs $L_i^V$ are related to the renormalized LECs $L_i^r(\mu)$ via~\cite{Ecker:1988te}
\begin{align}
	L_i^r(\mu)=\sum_{R=V,A,S,P}L_i^R+\hat{L}_i(\mu)\,,
\end{align}
where $L_i^R$ result from the contributions of the resonance $R$ (with $R=V$, $A$, $S$, and $P$ standing for the vector, axial-vector, scalar, and pseudo-scalar resonances, respectively), and $\hat{L}_i(\mu)$ represent the scale-dependent residue terms. The VMD picture suggests that $\hat{L}_i(M_\rho) \sim 0$. Using the numerical results of $L_i^r(M_\rho)$ collected in Table~3 of Ref.~\cite{Ecker:1988te}, together with the  inputs for $F_V$, $G_V$ and $F_{V_1}$ in Table~\ref{tab:input} of this work, we can then obtain
\begin{align}
|G_{V_1}|\simeq0.037~\text{GeV}\,.
\end{align}
The sign of $G_{V_1}$ is still unknown, and we shall assume it to be positive in this work, \textit{i.e.}, $G_{V_1}\simeq0.037~\text{GeV}$. As for $F_{V_n}^T$, the large-$N_C$ asymptotic analysis of the $\langle VV \rangle$, $\langle VT \rangle$ and $\langle TT \rangle$ correlators suggests that a pattern with possible alternation in sign,
\begin{align}
\xi_n=F_{V_n}^T/F_{V_n}=(-1)^n\frac{1}{\sqrt{2}}\,,
\end{align}
exists for the whole $J^{\mathcal{P}\mathcal{C}} = 1^{--}$ excited states~\cite{Cata:2008zc,Chen:2019vbr}. While $\xi_\rho$ is now confirmed to be positive~\cite{Becirevic:2003pn,Donnellan:2007xr,RBC-UKQCD:2008mhs,Cata:2009dq}, the sign of $\xi_{\rho^\prime}$ cannot be determined yet. For simplicity, we shall also assume a positive value of $\xi_{\rho^\prime}$ throughout this work.

Another task of this work is to estimate the resonance couplings $d_i$ present in the $\omega\pi$ form factors of Eqs.~\eqref{eq:V(s)}--\eqref{eq:T_3(s)}. Following Ref.~\cite{Guo:2008sh}, we obtain the couplings $d_3$, $d_s$, $d_M$, and $d_m$ by fitting the SM predicted $\tau^- \to \omega\pi^-\nu_\tau$ spectral function to the experimental data on the same observable~\cite{CLEO:1999heg}. To this end, assuming all these parameters to obey the normal distributions and following the general procedure of the method of least squares, we can obtain their best-fit values by minimizing the following $\chi^2$ function:
\begin{align}
\chi^2(\vec{\theta})=(\mathcal{O}_\mathrm{the.}(\vec{\theta})-\mathcal{O}_\mathrm{exp.})^\top \cdot \text{cov}^{-1} \cdot (\mathcal{O}_\mathrm{the.}
(\vec{\theta})-\mathcal{O}_\mathrm{exp.})\,,
\end{align}
where $\mathcal{O}_\mathrm{the.}$ and $\mathcal{O}_\mathrm{exp.}$ stand for the theoretical predictions and the experimental measurements of the spectral function, $\vec{\theta}=(d_3,\,d_s,\,d_m M_\pi^2+d_M M_\omega^2)$ denotes the vector of the (combination of) parameters to be constrained, and ``$\text{cov}$" represents the corresponding covariance matrix encoding the total uncertainties obtained by adding the experimental and theoretical ones in quadrature. Here we assume that all of the experimental inputs of the observables are independent of each other. The fitting results are given by\footnote{We thank Prof. Zhi-Hui Guo for providing us with the necessary numerical tables obtained in Ref.~\cite{CLEO:1999heg}. Note that our fitting results for these parameters are slightly different from that of Eq.~(36) in Ref.~\cite{Guo:2008sh}, because we have used a different input for $F_{V_1}$ during the fit.}
\begin{align}\label{eq:di}
	 d_3 =&-0.229 \pm 0.008\,, & d_s =&-0.259 \pm 0.039\,,& d_m M_\pi^2+d_M M_\omega^2=0.525\pm0.067\,,
\end{align} 
corresponding to $\chi^2_\mathrm{min}/\mathrm{d.o.f.}\simeq 2.0$. The remaining resonance couplings can be determined by following the same procedure as for $G_{V_1}$. After integrating out the vector resonances, the $\mathcal{O}(p^4)$ odd-intrinsic-parity R$\chi$T Lagrangian have contributions to the $\mathcal{O}(p^6)$ $\chi$PT Lagrangian~\cite{Bijnens:2001bb}. The proper matching between the two effective theories provides us with some useful constraints on the couplings of the $\mathcal{O}(p^4)$ odd-intrinsic-parity R$\chi$T Lagrangian. For convenience, we have listed the tedious expressions of these constraints in appendix~\ref{app:Birelations}. Solving Eqs.~\eqref{eq:B1}--\eqref{eq:B18}, together with the numerical values of $B_i$ ($i=1,\cdots,22$) taken from Ref.~\cite{Jiang:2015dba} and the results given in Eq.~\eqref{eq:di}, we can obtain the numerical results for all the required resonance parameters, which are all collected in Table~\ref{tab:couplings}. 

\begin{table}[t]
\centering
\tabcolsep 0.36in
\renewcommand\arraystretch{1.5}
\begin{tabular}{|l|l|l|}
\hline
   $c_1=-7.44\times10^{-3}$ & $d_1=6.08\times10^{-1}$  & $\kappa^V_1=-2.71\times10^{-3}$\\
   $c_2=-3.45\times10^{-3}$ & $d_2=-7.99\times10^{-2}$ & $\kappa^V_2=5.76\times10^{-3}$\\
   $c_3=1.86\times10^{-3}$  & $d_3=-2.29\times10^{-1}$ & $\kappa^V_3=8.81\times10^{-3}$\\
   $c_4=-9.59\times10^{-3}$ & $d_4=0$                  & $\kappa^V_4=-3.23\times10^{-2}$\\
   $c_5=3.99\times10^{-3}$  & $d_a=2.73$               & $\kappa^V_5=2.79\times10^{-2}$\\
   $c_6=2.16\times10^{-2}$  & $d_b=2.86$               & $\kappa^V_6=-6.99\times10^{-2}$\\
   $c_7=-3.40\times10^{-1}$ & $d_c=-1.29\times10^{-1}$ & $\kappa^V_7=-5.99\times10^{-2}$\\
                            & $d_d=-4.33\times10^{-1}$ & $\kappa^V_8=-2.60\times10^{-3}$\\
                            & $d_e=7.19\times10^{-1}$  & $\kappa^V_9=3.79\times10^{-3}$\\
                            & $d_f=7.49\times10^{-1}$  & $\kappa^V_{10}=3.53\times10^{-2}$\\
                            &                          & $\kappa^V_{18}=8.96\times10^{-3}$\\ 
\hline
\end{tabular}
\caption{Numerical results of the resonance couplings present in the $\mathcal{O}(p^4)$ odd-intrinsic-parity R$\chi$T Lagrangian, obtained by solving Eqs.~\eqref{eq:B1}--\eqref{eq:B18}, together with the numerical values of $B_i$ ($i=1,\cdots,22$) taken from Ref.~\cite{Jiang:2015dba} and the results given in Eq.~\eqref{eq:di}. See the texts for more details. \label{tab:couplings}}
\end{table}

\begin{figure}[t]
	\centering
	\includegraphics[width=0.478\textwidth]{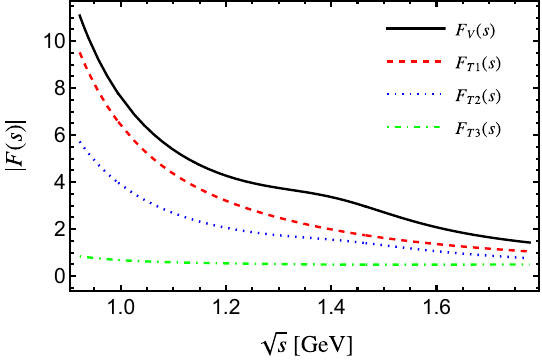}\;
	\includegraphics[width=0.498\textwidth]{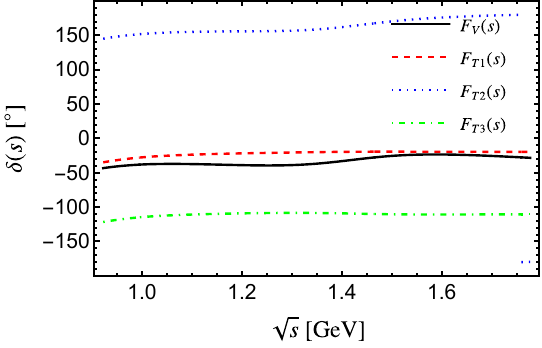}
	\caption{The moduli (left) and phases (right) of the form factors $F_V(s)$ (black solid), $F_{T1}(s)$ (red dashed), $F_{T2}(s)$ (blue dotted), and $F_{T3}(s)$ (green dot-dashed), obtained based on the analytic expressions given by Eqs.~\eqref{eq:FV}--\eqref{eq:FT3} and the input parameters taken from Tables \ref{tab:input} and \ref{tab:couplings}. \label{fig:FFs} } 
\end{figure}

Plugging into Eqs.~\eqref{eq:FV}--\eqref{eq:FT3} the numerical results of the resonance parameters mentioned above, we can finally obtain our numerical predictions for the $\omega\pi$ form factors as a function of the $\omega\pi$ invariant mass $\sqrt{s}$. As an illustration, we show in Figure~\ref{fig:FFs} the $\sqrt{s}$ dependence of both the moduli and the phases of the form factors $F_V(s)$ (black solid), $F_{T1}(s)$ (red dashed), $F_{T2}(s)$ (blue dotted), and $F_{T3}(s)$ (green dot-dashed). As can be seen from the left panel of Figure~\ref{fig:FFs}, there exists a slight bump lying around $\sqrt{s}\sim1.4~\text{GeV}$ in the vector form factor $F_V(s)$, indicating the existence of the resonance $\rho^\prime$. Although having also a contribution to the tensor form factors $F_{T1}(s)$, $F_{T2}(s)$, and $F_{T3}(s)$, the $\rho^\prime$ bump is not quite obvious, which attributes to the small weight of the $\rho^\prime$ contribution to these tensor form factors. On the other hand, the peak of the $\rho$ resonance is absent in all these form factors, because the $\omega\pi$ mass threshold is higher than the $\rho$ pole mass. As for the phases of these form factors, one can see from the right panel of Figure~\ref{fig:FFs} that they display only a small variation with respect to $\sqrt{s}$ in the whole $\omega\pi$ invariant mass range.

\subsection[Constraint on the tensor coefficient $\hat{\epsilon}_T$] {\boldmath Constraint on the tensor coefficient $\hat{\epsilon}_T$}

With the $\omega\pi$ form factors at hand, we can now estimate the NP impacts on the $\tau^- \to \omega \pi^- \nu_\tau$ decay. According to the analysis made in section~\ref{sec:EFT}, we know that only the non-standard tensor interaction can have a non-vanishing contribution to the decay, \textit{i.e.}, only a single NP parameter $\hat{\epsilon}_T$ is involved in our case. To set a constraint on $\hat{\epsilon}_T$, we can now make use of the measured branching ratio~\cite{ParticleDataGroup:2022pth},
\begin{align}
\mathcal{B}(\tau^- \to \omega \pi^- \nu_\tau)_\mathrm{exp}=(1.95\pm0.06)\%\,,
\end{align}
and the data on the spectral function $v(s)$ in 16 different bins taken from CLEO~\cite{CLEO:1999heg}. The corresponding theoretical expressions of these observables have already been provided in section~\ref{sec:EFT}. Applying again the method of least squares, we get the fitting result,
\begin{align}\label{eq:fit}
\hat{\epsilon}_T=(0.3\pm4.9)\times10^{-3}\,,
\end{align}
which corresponds to $\chi^2_\mathrm{min}/\mathrm{d.o.f.}\simeq1.7$. It can be seen that our fitting result of $\hat{\epsilon}_T$ is about one order of magnitude stronger than the bound obtained from a simultaneous fit to other one- and two-meson strangeness-conserving exclusive hadronic tau decays (explicitly, the data on $\tau^- \to \pi^- \nu_\tau$, $\tau^- \to \pi^- \pi^0 \nu_\tau$, and $\tau^- \to K^- K^0 \nu_\tau$ was considered)~\cite{Gonzalez-Solis:2020jlh}, $\epsilon_T=(-0.1\pm0.2^{+1.1\,+0.0}_{-1.4\,-0.1}\pm0.2)\times10^{-2}$, both bearing however larger uncertainties for the moment.

\subsection{NP impacts on the spectral function and forward-backward asymmetry}

\begin{figure}[t]
	\centering	
    \includegraphics[width=0.48\textwidth]{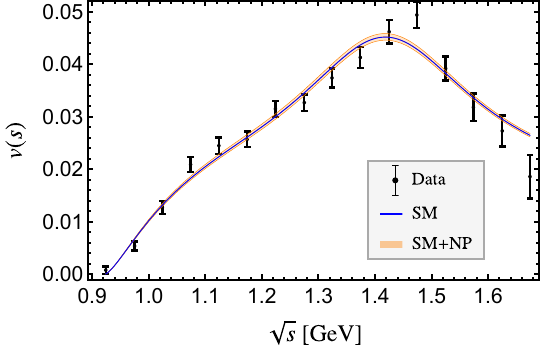}\;
    \includegraphics[width=0.49\textwidth]{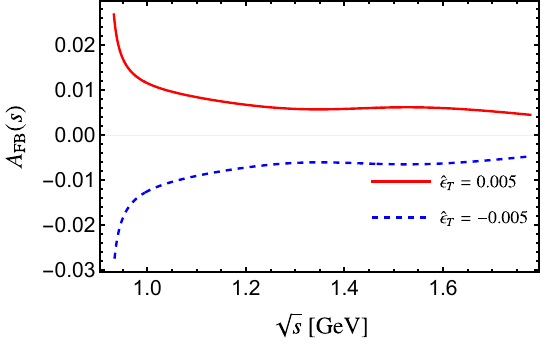}
	\caption{Left: the $\sqrt{s}$ distribution of the spectral function $v(s)$ with (orange band) and without (blue line) the NP contribution, where the bin data are taken from Ref.~\cite{CLEO:1999heg}. Right: the $\sqrt{s}$ distribution of the forward-backward asymmetry $A_\mathrm{FB}(s)$ predicted with two different values of the tensor coefficient, $\hat{\epsilon}_T=0.005$ (red solid) and $\hat{\epsilon}_T=-0.005$ (blue dashed).\label{fig:NPimpact} }
\end{figure}

To have an intuitive observation of the NP impacts on the $\tau^- \to \omega \pi^- \nu_\tau$ decay, we show in Figure~\ref{fig:NPimpact} the distributions of the spectral function $v(s)$ and the forward-backward asymmetry $A_\mathrm{FB}(s)$, both with and without the non-standard tensor contribution. As can be seen from the left panel of Figure~\ref{fig:NPimpact}, the NP contribution to the spectral function $v(s)$ is small compared to the SM prediction, and thus its presence provides very limited improvement on the fit. On the other side, since a non-zero forward-backward asymmetry $A_\mathrm{FB}(s)$ can arise only in the presence of a non-standard tensor contribution, we display the $\sqrt{s}$ distributions of this observable with two different inputs, \textit{i.e.}, $\hat{\epsilon}_T=0.005$ (red solid) and $\hat{\epsilon}_T=-0.005$ (blue dashed), which cover naively the allowed range of the tensor coefficient $\hat{\epsilon}_T$ given by Eq.~\eqref{eq:fit}. Here we would like to stress again that any measurement of $A_\mathrm{FB}(s)$ with a non-vanishing distribution could be served as a hint of the non-standard tensor interaction, and thus we strongly suggest further detailed studies of the observable at the Belle II experiment~\cite{Belle-II:2018jsg} as well as the proposed Tera-Z~\cite{CEPCStudyGroup:2018ghi,FCC:2018byv} and STCF~\cite{Achasov:2023gey} facilities.

\section{Conclusion}\label{sec:conclusion}

In this work, we have made a preliminary study of the NP effects in the $\tau^- \to \omega\pi^-\nu_\tau$ decay, which is usually employed as a sensitive test for the existence of the SCC. Without loss of generality, we started with the most general LEFT Lagrangian and found that, besides the SM vector interaction, only the non-standard tensor interaction can have a non-zero contribution to the decay, by comparing the $J^{\mathcal{P}G}$ decompositions of the quark currents and the $\omega\pi$ final state. As the current experimental upper limit on the SCC contribution is negligibly small, we have assumed the conservation of the $G$-parity in the decay and can, therefore, focus only on the FCC contribution both within the SM and in the NP scenario. 

To estimate the non-standard tensor impact on the $\tau^- \to \omega\pi^-\nu_\tau$ decay, a reliable calculation of the $\omega\pi$ tensor form factors is necessary. To this end, we have constructed for the first time the missing part of the R$\chi$T Lagrangian with external tensor sources, $\mathcal{L}_{VTP}$, which describes the interactions among the vector resonances, the lowest-lying pseudo-scalars, and the external tensor sources. The $\omega\pi$ vector and tensor form factors were then calculated in the R$\chi$T framework, with the relevant resonance couplings determined by combining the QCD short-distance constraints, the fit to the spectral function of $\tau^- \to \omega\pi^-\nu_\tau$ decay, as well as the matching between the $\mathcal{O}(p^4)$ odd-intrinsic-parity operators after integrating out the vector resonances and the $\mathcal{O}(p^6)$ $\chi$PT operators. With the obtained form factors as inputs, we subsequently made use of the measured branching ratio and the spectral function of the decay to set constraint on the tensor coefficient, $\hat{\epsilon}_T=(0.3\pm4.9)\times10^{-3}$, which is about one order of magnitude stronger than the previous bound obtained from a simultaneous fit to other one- and two-meson strangeness-conserving exclusive hadronic tau decays. The NP impacts on the distributions of the spectral function and the forward-backward asymmetry of $\tau^- \to \omega\pi^-\nu_\tau$ decay have also been investigated. We found that, while the non-standard tensor interaction can hardly affect the spectral function in a remarkable way, the forward-backward asymmetry is an ideal observable to probe this NP effect, since the latter arises only in the presence of a non-standard tensor contribution. Therefore, we suggest the Belle II experiment as well as the proposed Tera-Z and STCF facilities to make further detailed studies of the observable in the future.

\section*{Acknowledgements}

This work is supported by the National Natural Science Foundation of China under Grant Nos. 12135006 and 12075097, as well as the Fundamental Research Funds for the Central Universities under Grant No. CCNU22LJ004. FC is also supported by the Fundamental Research Funds for the Central Universities (11623330) and the 2024 Guangzhou Basic and Applied Basic Research Scheme Project for Maiden Voyage (2024A04J4190).

\appendix

\section{\boldmath Construction of $\mathcal{L}_{VTP}$} 
\label{app:constructVTP}

According to the basic ideas of chiral effective field theory~\cite{Weinberg:1978kz,Gasser:1983yg,Gasser:1984gg}, the chiral effective Lagrangian should satisfy all the symmetries of massless QCD, \textit{i.e.}, the Lorentz invariance, the parity ($\mathcal{P}$), the charge conjugation ($\mathcal{C}$), and the chiral symmetry, as well as the Hermiticity. To facilitate the construction of the chirally invariant Lagrangian, we choose the set of external fields that transform in the same manner under the chiral group and are endowed with a given chiral power counting~\cite{Bijnens:1999sh,Ecker:1988te,Cata:2007ns}. Specifically, for the external tensor fields, their transformation properties under the chiral group and the discrete symmetries are given, respectively, by~\cite{Cata:2007ns}
\begin{align}
  t_\pm^{\mu\nu}\to ht_\pm^{\mu\nu}h^\dagger\,, \qquad t^{\mu\nu}_\pm(\vec{x},t) \stackrel{\mathcal{P}}{\rightarrow} \pm t^{\mu\nu}_\pm(-\vec{x},t)\,, \qquad t^{\mu\nu}_\pm(\vec{x},t)\stackrel{\mathcal{C}}{\rightarrow} - t^{\mu\nu \top}_\pm(\vec{x},t)\,,
\end{align}
with $h \in SU(3)_V$. The chiral power counting for the external tensor fields is in principle arbitrary, but a convenient convention is to make their chiral dimension to coincide with that of the external scalar field~\cite{Cata:2007ns}
\begin{align}
  t_\pm^{\mu\nu} \sim \mathcal{O}(p^2).
\end{align}

\begin{table}[t]
	\tabcolsep 0.40in
	\renewcommand\arraystretch{1.5}
	\begin{center}
		\begin{tabular}{c c c c c}
			\hline
			\hline
			$\mathcal{O}$ & Dim & $\mathcal{P}$ & $\mathcal{C}$ & $\text{h.c.}$\\
			\hline
			$V^{\mu\nu}$ & 0 & $V_{\mu\nu}$ & $-(V^{\mu\nu})^\top$ & $V^{\mu\nu}$ \\
			$u^\mu$ & 1 & $-u^\mu$ & $(u^\mu)^\top$ & $u^\mu$ \\
			$\chi_{\pm}$ & 2 & $\pm \chi_{\pm}$ & $(\chi_{\pm})^\top$ & $\pm \chi_{\pm}$ \\
			$f_{\pm}^{\mu\nu}$ & 2 & $\pm f_{\pm \mu\nu}$ & $\mp (f_{\pm}^{\mu\nu})^\top$ & $f_{\pm}^{\mu\nu}$ \\
			$t_{\pm}^{\mu\nu}$ & 2 & $\pm t_{\pm \mu\nu}$ & $-(t_{\pm}^{\mu\nu})^\top$ & $\pm t_{\pm}^{\mu\nu}$ \\
			\hline
			\hline
		\end{tabular}
		\caption{Properties of the building blocks $V^{\mu\nu}$, $u^\mu$, $\chi_\pm$, $f_\pm^{\mu\nu}$, and $t_\pm^{\mu\nu}$ involved in the Lagrangians $\mathcal{L}_{VJP}$ and $\mathcal{L}_{VTP}$: chiral dimension (second column), parity (third column), charge conjugation (fourth column), and Hermitian conjugation (fifth column)~\cite{Cata:2007ns}. \label{tab:properties}}
	\end{center}
\end{table}

For convenience, we list in Table~\ref{tab:properties} the chiral dimension (Dim), the transformation properties under $\mathcal{P}$ and $\mathcal{C}$, as well as the Hermitian conjugation (h.c.) of the building blocks $V^{\mu\nu}$, $u^\mu$, $\chi_\pm$, $f_\pm^{\mu\nu}$, and $t_\pm^{\mu\nu}$. With these information at hand, we can then build all the operators that are invariant under the chiral, Lorentz, $\mathcal{P}$, and $\mathcal{C}$ transformations, up to a given chiral expansion order. Nevertheless, the resulting operators are not completely independent, and we have to make use of the Schouten identity, the equations of motion for the lowest-order $\chi$PT Lagrangian, and other relations about the derivatives, to reduce them to a minimal and irredundant basis. The explicit forms of these relations could be found in Refs.~\cite{Ruiz-Femenia:2003jdx,Cirigliano:2006hb,Jiang:2019hgs}. In this way, we obtain the following effective Lagrangian $\mathcal{L}_{VTP}$ with a complete and independent operator basis:
\begin{align}
	\mathcal{L}_{V T P}= & b_1 \epsilon_{\mu \nu \rho \sigma}\left\langle\left\{V^{\mu \nu}, t_{+}^{\rho \alpha}\right\} \nabla_\alpha u^\sigma\right\rangle +b_2 \epsilon_{\mu \nu \rho \sigma}\langle \{V^{\mu \alpha},t_{+}^{\nu\rho}\} \nabla_\alpha u^\sigma\rangle  \nonumber \\[1mm]
	&+i b_3 \epsilon_{\mu \nu \rho \sigma}\left\langle\left\{V^{\mu \nu}, t_{+}^{\rho \sigma}\right\} \chi_{-}\right\rangle +i b_4 \epsilon_{\mu \nu \rho \sigma}\left\langle\left\{V^{\mu \nu}, t_{-}^{\rho \sigma}\right\} \chi_{+}\right\rangle \nonumber \\[1mm]
	& +b_{5} \epsilon_{\mu \nu \rho \sigma}\left\langle\left\{\nabla_\alpha V^{\mu \nu}, t_{+}^{\rho \alpha}\right\} u^\sigma\right\rangle +b_{6} \epsilon_{\mu \nu \rho \sigma}\left\langle\left\{\nabla_\alpha V^{\mu \alpha}, t_{+}^{\nu\rho}\right\} u^\sigma\right\rangle \nonumber \\[1mm]
	&+b_{7} \epsilon_{\mu \nu \rho \sigma}\left\langle\left\{\nabla^\mu V^{\nu \rho}, t_{+}^{\sigma \alpha}\right\} u_\alpha\right\rangle +i b_8 \epsilon_{\mu \nu \rho \sigma}\left\langle V^{\mu \nu}t_{-}^{\rho \sigma}\right\rangle\langle \chi_{+} \rangle \nonumber \\[1mm]
	&+b_9\epsilon_{\mu \nu \rho \sigma}\langle V^{\mu \nu} \nabla_\alpha u^\rho\rangle\langle t_{+}^{\sigma \alpha} \rangle +b_{10} \epsilon_{\mu \nu \rho \sigma}\langle V^{\mu\nu}\nabla^{\rho}u_\alpha \rangle\langle  t_{+}^{\sigma\alpha} \rangle \nonumber \\[1mm]
	&+i b_{11} \epsilon_{\mu \nu \rho \sigma}\left\langle V^{\mu \nu}  \chi_{-}\right\rangle\langle t_{+}^{\rho \sigma}\rangle +i b_{12} \epsilon_{\mu \nu \rho \sigma}\left\langle V^{\mu \nu}\chi_{+}\right\rangle\langle t_{-}^{\rho \sigma} \rangle \nonumber \\[1mm]
	& + b_{13} \epsilon_{\mu \nu \rho \sigma}\left\langle\nabla_\alpha V^{\mu \nu} u^\rho\right\rangle\langle t_{+}^{\sigma \alpha} \rangle +b_{14} \epsilon_{\mu \nu \rho \sigma}\left\langle\nabla_\alpha V^{\mu \alpha}u^\nu\right\rangle\langle t_{+}^{\rho \sigma}\rangle \nonumber\\[1mm]
	& +b_{15} \epsilon_{\mu \nu \rho \sigma}g_{\alpha\beta}\left\langle\nabla^\mu V^{ \nu\alpha} u^\rho\right\rangle\langle t_{+}^{\sigma \beta} \rangle\,. \label{eq:vtpL}
\end{align}
It can be seen that the structures of the first seven terms in $\mathcal{L}_{VTP}$ agree with that of the effective Lagrangian $\mathcal{L}_{VJP}$ given by Eq.(\ref{eq:VJP}). This is understandable because the two building blocks $t_{\pm}^{\mu\nu}$ and  $f_{\pm}^{\mu\nu}$ have the same chiral dimension and transform in the same manner under the chiral group. Furthermore, since $t_+^{\mu\nu}$ has also the same $\mathcal{P}$ and $\mathcal{C}$ transformation properties as that of $f_{+}^{\mu\nu}$, the six terms involving $t_+^{\mu\nu}$ in $\mathcal{L}_{VTP}$ have a one-to-one correspondence to that of $\mathcal{L}_{VJP}$. However, as the external tensor fields $t_{\pm}^{\mu\nu}$ have a non-vanishing trace in flavor space, we have to include the operators involving the traces over $t_{\pm}^{\mu\nu}$ in the Lagrangian $\mathcal{L}_{VTP}$, as shown by the remaining terms in Eq.~\eqref{eq:vtpL}. With respect to the operators involving the building blocks $f_-^{\mu\nu}$ and $t_-^{\mu\nu}$, on the other hand, because of their obviously different transformation properties under the charge conjugation and Hermitian conjugation, the forms of these operators can be different from each other.

Although the Lagrangian constructed above contains the Levi-Civita tensor, it still belongs to the even-intrinsic-parity sector and is of $\mathcal{O}(p^4)$. This is different from the odd-intrinsic-parity sector of the $\chi$PT Lagrangian, where the lowest-order odd-intrinsic-parity operators involving the external tensor sources should be of $\mathcal{O}(p^8)$, as demonstrated in Ref.~\cite{Cata:2007ns}. Actually, we can bring all possible contractions of the indices of the tensor sources with that of the Levi-Civita tensors into the forms without the Levi-Civita tensors, in terms of the following relations~\cite{Cata:2007ns}: 
\begin{align}\label{eq:tenre}
 \epsilon_{\mu\nu\rho\sigma}t_{\pm}^{\rho\sigma}X^{\mu\nu}=&2it_{\mp\mu\nu}X^{\mu\nu}\,,\nonumber\\[0.2cm]
 \epsilon_{\mu\nu\rho\sigma}t_{\pm}^{\mu\alpha}{X_{\alpha}}^{\nu\rho\sigma}=&3it_{\mp\rho\sigma}{X_{\nu}}^{\nu\rho\sigma}\,, 
\end{align}
where $X$ stands for any generic chiral tensor made out of the building blocks for constructing the chiral effective Lagrangian. In this way, another equivalent form of Eq.~\eqref{eq:vtpL} can be obtained, from which the even-intrinsic-parity property of $\mathcal{L}_{VTP}$ can be clearly seen.

\section{\boldmath Matching between the odd-intrinsic-parity operators of $\mathcal{O}(p^4)$ in R$\chi$T and that of $\mathcal{O}(p^6)$ in $\chi$PT} \label{app:Birelations}

Integrating out the vector resonances in the full set of the $\mathcal{O}(p^4)$ odd-intrinsic-parity R$\chi$T Lagrangian (in the absence of NP, \textit{i.e.}, without $\mathcal{L}_{V T P}$) will automatically introduce contributions to the $\mathcal{O}(p^6)$ $\chi$PT Lagrangian~\cite{Ecker:1988te,Cirigliano:2006hb,Kampf:2011ty}. Considering the contributions from the first and the second nonet of vector resonances, we obtain the following matching relations:
\begin{align}
	B_{1}=&-\frac{G_{V}^2}{4M_{V}^4}(2d_1+16d_2-3d_3)-\frac{G_{V}G_{V_1}}{4M_{V}^2 M_{V_1}^2}(2d_a+2d_b-3d_c+5d_d+8d_f) \nonumber \\[1mm] 
    &+\frac{\sqrt{2}G_V}{M_V^3}(\kappa^V_3-2\kappa^V_9+\kappa^V_{10})\,, \label{eq:B1}\\[2mm]
	B_{2}=&-\frac{1}{2\sqrt{2} M_V^3}\left[ 4 G_{V} c_4 -F_V \kappa^V_4 \right]\,, \label{eq:B2} \\[2mm]
	B_{4}=&-\frac{F_V G_V}{24M_V^4}(6d_1+48d_2+d_3)\nonumber \\[1mm]  
    &-\frac{1}{24 M_V^2 M_{V_1}^2} \left[ F_{V} G_{V_1}(6d_a-7d_c+24d_f)+G_{V}F_{V_1}(6d_b+d_d+24d_f) \right] \nonumber \\[1mm] 
	&+\frac{1}{12\sqrt{2}M_V^3}\left[ G_V(6c_2+24c_3+c_6)+4F_V(3\kappa^V_1+2\kappa^V_5+\kappa^V_6+\kappa^V_8-3\kappa^V_9) \right]\,,  \\[2mm]
	B_{5}=&-\frac{F_V G_V}{4 M_V^4}d_3-\frac{1}{4 M_V^2 M_{V_1}^2}\left[ G_V F_{V_1}(2d_c-d_d)+ F_V G_{V_1}(d_c-4 d_d) \right] \nonumber \\[1mm]  
    &+\frac{1}{2\sqrt{2}M^{3}_V}\left[ G_V(2c_5-c_6)+ 2 F_V \kappa^V_{10} \right]\,,  \\[2mm]
	B_{6}=&\frac{F_VG_V}{6nM_V^4}(3d_1-d_3)\nonumber \\[1mm]  
    &+\frac{1}{6nM_V^2M_{V_1}^2}\left[ F_{V}G_{V_1}(3d_a-5d_c+6d_d)+ G_{V}F_{V_1}(3d_b-3d_c+2d_d) \right]  \nonumber\\[1mm] 
	&-\frac{1}{3n\sqrt{2}M^{3}_V}\left[ G_V(3c_2-3c_5+2c_6) +F_V(6\kappa^V_1+4\kappa^V_5+2\kappa^V_6+2\kappa^V_8+3n \kappa^V_{18}) \right]\,, \\[2mm]
	B_{7}=&\frac{F_V^2}{8M_V^4}(d_1+8d_2-d_3)+\frac{F_VF_{V_1}}{8M_V^2M_{V_1}^2}(d_a+d_b-d_c+8d_f)\nonumber \\[1mm]  
    &-\frac{F_V}{4\sqrt{2}M_V^3}(c_1+c_2+8c_3-c_5)\,, \\[2mm]
	B_{8}=&-\frac{F_V^2}{8nM_V^4}(d_1-d_3)-\frac{F_VF_{V_1}}{8nM_V^2M_{V_1}^2}(d_a+d_b-d_c)+\frac{F_V}{4n\sqrt{2}M_V^3}(c_1+c_2-c_5)\,, \\[2mm]
	B_{11}=&-\frac{F_V}{2\sqrt{2}M_V^3}(2c_4-\kappa^V_4)\,, \label{eq:B11}\\[2mm]
	B_{12}=&\frac{G_V^2}{M_V^4}d_3 +\frac{G_VG_{V_1}}{M_V^2M_{V_1}^2}(d_c-d_d)+\frac{\sqrt{2}G_V}{M_V^3}(\kappa^V_1-\kappa^V_2-\kappa^V_3)\,,\\[2mm]
	B_{13}=&\frac{7F_VG_V}{6M_V^4}d_3\nonumber \\[1mm]  
    &+\frac{1}{6M_V^2M_{V_1}^2}\left[ G_VF_{V_1}(3d_a-3d_b+3d_c+4d_d)-F_{V}G_{V_1}(3d_a-3d_b+d_c+6d_d) \right] \nonumber\\[1mm] 
	&-\frac{1}{3\sqrt{2}M_V^3}\left[ G_V(3c_1-3c_2+3c_5+4c_6)-2 F_V(3\kappa^V_1+2\kappa^V_5+\kappa^V_6+\kappa^V_8) \right]\,, \\[2mm]
	B_{14}=&-\frac{F_V G_V}{6 M_V^4}d_3+\frac{1}{6 M_V^2 M_{V_1}^2}\left[ F_V G_{V_1} d_c-G_VF_{V_1} d_d  \right]\nonumber \\[1mm] 
    &+\frac{1}{3\sqrt{2}M_V^3}\left[ G_V c_6+F_V(3\kappa^V_1-3\kappa^V_3+2\kappa^V_5+\kappa^V_6+\kappa^V_8) \right]\,, \\[2mm]
	B_{15}=&\frac{5F_VG_V}{6M_V^4}d_3-\frac{1}{6M_V^2M_{V_1}^2}\left[ G_VF_{V_1}(3d_c-2d_d)+F_V G_{V_1}(2d_c-3d_d) \right]  \nonumber \\[1mm] 
	&+\frac{1}{3\sqrt{2}M_V^3}\left[ G_V(3c_5-2c_6)-F_V (3\kappa^V_2-2\kappa^V_5+\kappa^V_6-\kappa^V_8) \right]\,,  \\[2mm]
	B_{16}=&-\frac{G_V^2}{M_V^4}d_3-\frac{G_{V}G_{V_1}}{M_V^2M_{V_1}^2}(d_c-d_d)-\frac{\sqrt{2}G_V}{M_{V}^3}(\kappa^V_6-\kappa^V_7-\kappa^V_8)\,,\\[2mm]
	B_{17}=&\frac{1}{12M_V^4}\left[ F_VG_V(d_3-3d_4)-6 G_V^2 d_4 \right]\nonumber \\[1mm] &-\frac{1}{12 M_V^2M_{V_1}^2}\left[ F_VG_{V_1}(d_c+3d_e)-G_V F_{V_1} d_d+6 G_VG_{V_1} d_e \right]  \nonumber \\[1mm] 
	&-\frac{1}{6\sqrt{2} M_V^3}\left[ G_V(c_6-12\kappa^V_5)-2F_V(2\kappa^V_5+\kappa^V_6+\kappa^V_8) \right]\,,  \\[2mm]
	B_{19}=&-\frac{1}{2M_{V}^4}\left[ 2 F_VG_V(d_1-d_4)+F_V^2 d_4 \right] \nonumber \\[1mm] &-\frac{1}{2M_V^2M_{V_1}^2}\left[ G_VF_{V_1}(d_a+d_b-d_c-2d_e)+F_V G_{V_1}(d_a+d_b-d_c+2d_d)+F_V F_{V_1} d_e \right]  \nonumber\\[1mm] 
	&-\frac{1}{\sqrt{2}M_V^3}\left[ G_V(c_1-c_2+c_5+2c_7)-F_V(c_7-2\kappa^V_6) \right]\,,  \\[2mm]
	B_{20}=&\frac{1}{12 M_V^4}\left[ 2F_VG_V(d_3-3d_4) -3 F_V^2 d_4 \right]\nonumber \\[1mm] &-\frac{1}{12 M_V^2 M_{V_1}^2}\left[ 2F_V G_{V_1} d_c -2G_V F_{V_1}(d_d+3d_e) + 3F_V F_{V_1} d_e \right] \nonumber \\[1mm] 
	&-\frac{1}{6\sqrt{2}M_V^3}\left[ 2G_V(c_6+3c_7) - F_V (3c_7+20\kappa^V_5+4\kappa^V_6+4\kappa^V_8) \right]\,,  \\[2mm]	
	B_{21}=&-\frac{1}{12M_V^4}\left[ 2F_VG_V(2d_3-3d_4)-3F_V^2 d_4 \right]\nonumber \\[1mm]  &-\frac{1}{12M_V^2M_{V_1}^2}\left[ 2G_{V}F_{V_1}(3d_c-d_d-3d_e)+2F_{V}G_{V_1} (d_c-3d_d) -3 F_V F_{V_1} d_e \right] \nonumber \\[1mm] 
	&+\frac{1}{6\sqrt{2}M_V^3}\left[ 2G_V(3c_5-c_6+3c_7) -F_V (3c_7+4\kappa^V_5+2\kappa^V_6-6\kappa^V_7+2\kappa^V_8) \right]\,,  \\[2mm]
	B_{22}=&-\frac{F_V^2}{2M_V^4}d_3-\frac{F_VF_{V_1}}{2M_V^2M_{V_1}^2}(d_c-d_d)+\frac{F_V}{\sqrt{2}M_V^3}(c_5-c_6)\,,\label{eq:B18}
\end{align}
where $n=3$ is the number of flavors, and $c_i$, $d_i$, and $\kappa^V_i$ denote the resonance couplings of the $\mathcal{O}(p^4)$ odd-intrinsic-parity R$\chi$T Lagrangian, whereas $B_i$ are the LECs of the $\mathcal{O}(p^6)$ $\chi$PT Lagrangian~\cite{Kampf:2011ty}. 

\bibliographystyle{JHEP}
\bibliography{reference}

\end{document}